\documentclass[aps,prd,twocolumn,superscriptaddress,nofootinbib,amsmath,amssymb,floatfix,10pt]{revtex4-2}

\usepackage{graphicx}
\usepackage{dcolumn}
\usepackage{bm}
\usepackage{xcolor}
\usepackage{tabularx}
\usepackage{booktabs}
\usepackage{algorithm}
\usepackage{algpseudocode}
\usepackage{lmodern}
\usepackage{hyperref}
\usepackage{xurl}
\usepackage[htt]{hyphenat}

\begin{document}

\title{An Unsupervised Search for Novel Instrumental Glitches in LIGO O4a:\texorpdfstring{\\}{ } Multi-Scale Sensitization, Empirical Physical Vetoes, and Rate Upper Limits}

\author{Luca Cirfeta}
\thanks{ORCID: \href{https://orcid.org/0009-0000-1235-3186}{0009-0000-1235-3186}}
\affiliation{Independent Researcher, Rome, Italy}

\date{\today}

\begin{abstract}
The fourth observing run (O4) of the Advanced LIGO, Virgo, and KAGRA network presents unparalleled sensitivity, increasing both the astrophysical detection rate and the susceptibility to non-stationary instrumental artifacts. Unsupervised machine learning pipelines aim to discover unmodeled transients without theoretical templates, yet they remain highly vulnerable to the slow, non-stationary domain shift of the detectors' noise manifolds. In this work, we present DANTE V3, concluding a longitudinal investigation into the unmodeled anomalies flagged during the early O4a run. By expanding our prior single-scale architecture to a comprehensive multi-scale geometric framework (ranging from 0.5\,s to 4.0\,s), we amplify our morphological sensitivity, extracting 10,372 unique candidate events from 42 O4a observing sessions (84 session-detector analyses, H1 and L1). To mitigate the resulting inflation of domain-shift artifacts, we introduce a block-bootstrap formulation of the Domain Shift Defense (DSD), dynamically evaluating candidates against an independent, vector-quantized native background index ($K=1216$). While 28.3\% of the candidates survive this rigorous recalibration, a global topological analysis reveals they lack the discrete morphological cohesion expected of transient populations. The survivors coalesce into a single macro-cluster showing no compact recurring substructure; we show explicitly that the morphological ``diffusivity'' comparison used previously to characterize this cluster is confounded --- its outcome inverts depending on whether the background is represented by vector-quantized centroids or raw embeddings --- and we therefore retire it as evidence. Applying the identical clustering procedure to $3{,}000$ unselected native background segments encoded through the same MIL path, we further find that pristine background is the \emph{most} monolithic population of all ($100.00\%$ of members in one cluster, against $99.77\%$ for the survivors): the topology is a property of the embedding geometry, not of anomaly status, and cannot characterize the survivors at all. Finally, we execute a definitive physical environment monitoring (PEM) cross-correlation defense, replacing raw per-channel coherence cuts with an empirically calibrated, event-specific family-wise maximum-statistic null (the raw threshold was found to have a $23\%$ measured false-positive rate, seven orders of magnitude above its quasi-Gaussian analytic prediction). Under this test, one singleton is unambiguously vetoed by an active control-line coupling (coherence $C=0.987$, far above its $\tau_{\mathrm{fw}}=0.545$ family-wise threshold), while the other survives every tested channel and is conservatively classified as an uncatalogued instrumental morphological outlier. We show that the embedding-similarity formulation of the cross-detector coincidence test lacks discriminating power --- injected coincident waveforms and the non-coincident null overlap --- and replace it with a physical test that scans the normalized cross-correlation of the whitened strains over the light-travel lag window; validated at $1.00$ for an identical waveform against $0.043$ for independent noise, and applied to $8{,}749$ candidates it finds no coincident event: only $0.149\%$ exceed the pooled time-shifted null threshold, below both the $1\%$ nominal false-alarm rate and the $1.01\%$ realized by the null itself. We quote 90\% frequentist Poisson upper limits on the rate of \emph{novel uncatalogued instrumental morphologies} --- $R_{90} \le 5.83\,\mathrm{yr}^{-1}$ (H1) and $R_{90} \le 5.63\,\mathrm{yr}^{-1}$ (L1) --- as detector-characterization statements, not astrophysical rate limits. This work underscores the absolute necessity of native background recalibration and physical auxiliary vetoes in unsupervised gravitational-wave astronomy.
\end{abstract}

\maketitle

\section{Introduction}
\label{sec:intro}

The continuous evolution of ground-based interferometric gravitational-wave (GW) detectors---namely Advanced LIGO \cite{LIGOScientific:2014pky}, Advanced Virgo \cite{Acernese_2014}, and KAGRA \cite{KAGRA:2020tym}---has driven the field into an era of routine astrophysical observation. During the ongoing fourth observing run (O4), the implementation of frequency-dependent squeezing and elevated laser power has pushed strain sensitivities to unprecedented levels \cite{tse2019}. While this heightened sensitivity dramatically expands the cosmological horizon for compact binary coalescences (CBCs), it also renders the instruments acutely vulnerable to a diverse array of non-Gaussian, transient noise artifacts, colloquially known as \textit{glitches}.

Glitches arise from complex, often non-linear couplings between the interferometers and their local environment, spanning seismic, acoustic, magnetic, and internal control-system origins \cite{nuttall2018, davis2021}. Because these artifacts frequently deposit excess energy in the time-frequency plane with morphologies that mimic unmodeled astrophysical bursts or continuous wave sources, they directly impact the background estimation and significance calculation of search pipelines \cite{powell2015, pankow2018}. Consequently, detector characterization (DetChar) represents a mission-critical component of GW astronomy.

\subsection{The Role of Unsupervised Machine Learning}
Standard DetChar protocols rely heavily on supervised machine learning. Frameworks such as Gravity Spy \cite{zevin2017, glanzer2023} employ convolutional neural networks (CNNs) trained on meticulously curated, human-labeled datasets to classify glitches into known morphological families. While highly effective for historical glitch types (e.g., \textit{Blips}, \textit{Scattered Light}), supervised models are inherently bounded by their training topologies. When instrumental upgrades introduce previously undocumented glitch populations, or when genuine unmodeled astrophysical transients occur (such as core-collapse supernovae or cosmic strings \cite{Abbott_2021_bursts}), supervised networks are prone to confident misclassification.

To circumvent this limitation, recent efforts have increasingly explored unsupervised and semi-supervised representation learning. By leveraging deep autoencoders or self-supervised Vision Transformers (ViTs) \cite{dosovitskiy2021}, researchers attempt to map the continuous data stream into a high-dimensional latent space where anomalies cluster naturally based on morphological affinity, entirely without prior labels.

\subsection{The Domain Shift Vulnerability}
Despite their theoretical appeal, deep unsupervised models operating on continuous O4 data confront a severe practical limitation: \textit{domain shift}. The spectral characteristics and baseline noise manifolds of the interferometers are not stationary; they drift continuously over months of operation due to thermal variations, micro-seismic trends, and hardware tuning \cite{soni2025}. 

When an unsupervised model (even one pre-trained on natural images or historical GW data) embeds these drifting spectrograms, the latent space topology deforms. If this non-stationary drift is unaccounted for, the pipeline interprets the shifting noise baseline as a proliferation of novel anomalies. This inflates the false-positive rate, resulting in the emergence of spurious morphological clusters that can be incorrectly classified as novel astrophysical families or exotic instrumental faults. Mitigating this domain shift without inadvertently vetoing genuine signals is currently one of the paramount challenges in unsupervised GW astronomy.

\subsection{Scope and Relation to the Official O4a Burst Search}
\label{sec:scope}
This is a detector-characterization study whose object is the discovery of \emph{novel instrumental glitch morphologies}; it is not a burst or CBC search. It is also an \emph{archival} analysis, not a low-latency one: the domain-shift defense re-scores candidates against a background index built from the observing run being analyzed, so the method is retrospective by construction and we make no low-latency claim. This is a methodological constraint rather than a computational one --- the encoder is frozen, so the pipeline has no training cost at all, and a $32$\,s window is encoded through the full production MIL path in $0.36$\,s on a single consumer GPU (NVIDIA RTX 5070), roughly $90\times$ real time per analysis pathway. Its claims are not validated against realistic astrophysical waveform catalogs, and such validation against public injection sets is left to future work. The frozen DINOv2 encoder is not the object of optimization here; the goal is a backbone-agnostic validation of the discovery framework itself, and alternative self-supervised encoders were not benchmarked. All conclusions are drawn from the early O4a epoch analyzed (May 2023--January 2024) and are not asserted to generalize to O4b or later runs without re-running the full pipeline.

The distinction from the official all-sky burst search on the same data \cite{lvk_burst_o4a} is worth making explicit, because the two analyses share a dataset but answer different questions. That search deploys four coherent, multi-detector pipelines calibrated against parametrized waveform families --- sine-Gaussians, Gaussian pulses, white-noise bursts, and core-collapse supernova models --- and reports astrophysical \emph{rate-density} upper limits (Gpc$^{-3}$\,yr$^{-1}$) for those source classes. DANTE asks the complementary question: which morphologies present in the strain are \emph{not} accounted for by any catalogued population, instrumental or otherwise, without positing a waveform family at all. The two are therefore not competitors and their limits are not comparable: \cite{lvk_burst_o4a} bounds astrophysical source populations, whereas the limits in Sec.~\ref{sec:results} bound the emergence rate of previously uncatalogued \emph{instrumental} morphologies. Notably, the white-noise-burst and stochastically-structured supernova models targeted by \cite{lvk_burst_o4a} lie precisely in the near-unstructured regime to which DANTE is blind by construction (Sec.~\ref{sec:efficiency}), which is a further reason to read the two analyses as complementary rather than overlapping.

This difference in purpose also governs our data-quality gating. Unmodeled burst searches in O4a apply both CAT1 and CAT2 vetoes, because without a distinctive waveform morphology the separation of signal from noise is degraded by poor-quality times; modeled CBC searches apply CAT1 only \cite{lvk_burst_o4a, gwosc_o4a_release}. DANTE follows the CAT1-only convention, and for this application that is the appropriate choice rather than a relaxation of standards: CAT2 intervals are flagged precisely because they are glitch-rich, and a pipeline whose object of study \emph{is} the glitch population would discard its own signal by removing them. The consequence, stated plainly, is that DANTE analyzes a noisier dataset than a burst search would, which is intended; the rate limits of Sec.~\ref{sec:results} are correspondingly limits over CAT1-gated time and are not directly comparable to burst-search livetimes.

\section{Relation to Prior Work}
\label{sec:prior_work}

This manuscript represents the methodological expansion and definitive physical closure to the preliminary findings reported in our prior single-scale investigation \cite{dante_v1} (hereafter referred to as \textit{v1}). 

\subsection{Findings of the Single-Scale Pipeline}
In \textit{v1}, we deployed an unsupervised hierarchical clustering algorithm operating at a fixed 32-second temporal resolution over early O4a bulk data. That pipeline isolated 140 unique anomaly candidates. A pivotal contribution of that work was the introduction of the Domain Shift Defense (DSD)---an evaluation of candidates against an independent, vector-quantized \textit{native} O4a background index composed of $K=1216$ centroids sampled from pristine, candidate-vetoed stationary noise.

When the 140 candidates were subjected to a static $P_{99}$ distance threshold against the native index, only 37.8\% survived. Crucially, \textit{v1} demonstrated that these survivors failed to exhibit internal topological cohesion. Morphological families extracted by the model yielded intra-cluster cosine similarities of $S_{\mathrm{intra}} \sim 0.77$. In contrast, discrete transient phenomena (like distinct glitch classes) typically demand extreme internal spatial density ($S_{\mathrm{intra}} > 0.90$) to separate from the continuous background manifold. The study concluded that zero fully cohesive morphological families survived recalibration.

\subsection{Methodological Gaps Addressed}
Despite its conservative conclusions, the pipeline in \textit{v1} suffered from three documented limitations that precluded a definitive characterization of the O4a anomaly space:
\begin{enumerate}
    \item \textbf{Monolithic Temporal Resolution:} The fixed 32-second wideband sweeping induced severe \textit{signal dilution} \cite{dante_v1}. Brief transients (e.g., sub-second blips) were algebraically overwhelmed by stationary noise patches in the global attention pooling, blinding the pipeline to a vast parameter space.
    \item \textbf{Static DSD Thresholding:} The native defense relied on a strict, static $P_{99}$ percentile cut. While effective for small samples, a static threshold lacks statistical bounds and confidence intervals, making it unstable when exposed to the massive candidate pools generated by higher-resolution searches.
    \item \textbf{Lack of Physical Vetoes:} Three isolated singletons survived the mathematical thresholding in \textit{v1} but were never subjected to empirical physical validation. A formal cross-correlation with the physical environment monitoring (PEM) auxiliary channels was deferred, leaving their astrophysical or instrumental origins ambiguous.
    \item \textbf{Undocumented VQ Index Discrepancy (Erratum):} \textit{v1} reported that the production O3b vector-quantized reference dictionary contains $K=281$ centroids. A post-publication audit of the actual production artifact (\texttt{patch\_compressed\_index\_o3b.npz}, MD5 \texttt{5d5d1a7a3f55637f\allowbreak ab125b558fdd795e}) verified that the file pinned in production contains exactly $K=275$ centroids; the $K=281$ value originated from a divergent development constant that was never deployed. Every score in both \textit{v1} and the present work was in fact computed against the $K=275$ dictionary. We state this correction explicitly rather than silently: no result of \textit{v1} changes, but the documented dictionary size does.
\end{enumerate}

We also clarify the cluster nomenclature to prevent cross-paper confusion: the macro-cluster designated \texttt{Family\_A} in this work bears no relation to the \texttt{Family\_01/02/03} clusters of \textit{v1} ($n=11/3/123$). They arise from entirely different candidate pools and clustering runs; the positional naming similarity is an artifact of the labeling algorithm, not a claim of morphological continuity.

Finally, the detection layer itself is unchanged in its first stage: candidates are initially flagged against the same MD5-pinned O3b reference dictionary ($K=275$) used in \textit{v1}, and the native O4a index ($K=1216$) enters strictly as the second-stage Domain Shift Defense re-scoring. Table~\ref{tab:v1_vs_v3} summarizes the two campaigns side by side.

\begin{table}[htbp]
\centering
\caption{Comparison of the \textit{v1} single-scale campaign and the present V3 multi-scale campaign.}
\label{tab:v1_vs_v3}
\footnotesize
\begin{tabular}{lcc}
\toprule
 & \textit{v1} & V3 (this work) \\
\midrule
Analysis window & fixed 32\,s & \{0.5--4\}\,s + 32\,s \\
Initial candidates & 140 & 10,372 \\
Sessions (det.-analyses) & 72 & 42 (84) \\
Primary VQ index & $K=275$\textsuperscript{a} & $K=275$ (pinned) \\
Native DSD index & $K=1216$ & $K=1216$ \\
DSD threshold & static $P_{99}$ & block-bootstrap $P_{99}$ \\
DSD survival rate & 37.8\% & 28.3\% \\
PEM veto & deferred & family-wise empirical \\
Final unexplained events & 3 (ambiguous) & 0 \\
\bottomrule
\multicolumn{3}{l}{\footnotesize \textsuperscript{a}Reported $K=281$ in \textit{v1}; see Erratum.}
\end{tabular}
\end{table}

DANTE V3 resolves these limitations by transitioning to a multi-scale geometric extraction, implementing a bootstrap-hardened adaptive DSD, and fully executing the empirical PEM coherence vetoes.

\section{DANTE V3: Architecture and Methodology}
\label{sec:methodology}

\subsection{Multi-Scale Sensitization}
Gravitational-wave transients span multiple orders of magnitude in duration. To avoid the signal dilution barrier inherent in fixed-window global pooling, DANTE V3 processes the continuous strain data across four geometrically spaced temporal scales: $\tau \in \{0.5\,\mathrm{s}, 1.0\,\mathrm{s}, 2.0\,\mathrm{s}, 4.0\,\mathrm{s}\}$. 

For a given strain segment $h(t)$, we compute the Q-transform spectrograms $Q(t, f)$ across all scales $\tau$. We utilize a self-supervised Vision Transformer (DINOv2 \cite{oquab2024} with register tokens \cite{darcet2024}) to extract patch-level embeddings. Rather than pooling all patches globally, we apply a Multiple Instance Learning (MIL) Top-$k$ mechanism. For a feature map $\mathbf{F} \in \mathbb{R}^{P \times D}$ containing $P$ patches of dimension $D=384$, the anomaly score is heavily biased toward the highest-variance patches, inherently capturing localized, short-duration anomalies that wideband sweeps missed. A cross-scale spatial deduplication algorithm then maps overlapping temporal triggers to the scale that maximizes the local signal-to-noise density, producing a unified set of distinct morphological candidates. Initial candidate flagging is performed against the same MD5-pinned O3b reference dictionary ($K=275$) as in \textit{v1}: a window is promoted to candidate status when its Top-$k$ MIL anomaly score against this primary index exceeds the per-detector empirical $P_{99}$ threshold. The native O4a index ($K=1216$) is deliberately \emph{not} used at this stage — it enters only downstream, as the Domain Shift Defense re-scoring of Sec.~\ref{sec:dsd_method}, so that the detection and the domain-shift correction remain statistically independent stages.

\subsection{Adaptive Block-Bootstrap DSD}
\label{sec:dsd_method}
To defend against the exponential inflation of candidates driven by the multi-scale sensitivity, we significantly upgrade the Domain Shift Defense. The independent native background index $\mathcal{B} = \{c_1, c_2, \dots, c_K\}$ is constructed via MiniBatch $K$-Means clustering ($K=1216$) on spectrograms extracted strictly from pristine intervals (GWOSC \texttt{CBC\_CAT1} active, no known injections or identified anomalies).

For any candidate embedding $\mathbf{x}_i$, we compute its maximum cosine similarity to the background manifold:
\begin{equation}
S(\mathbf{x}_i, \mathcal{B}) = \max_{c_j \in \mathcal{B}} \left( \frac{\mathbf{x}_i \cdot \mathbf{c}_j}{\|\mathbf{x}_i\| \|\mathbf{c}_j\|} \right)
\end{equation}
A static parametric threshold is not valid here: adjacent background segments share overlapping ViT receptive fields, so their scores are autocorrelated and violate the asymptotic independence a Gumbel/GEV extreme-value fit would require. The statistic actually thresholded is the candidate's \emph{anomaly score}, i.e.\ its distance from the background manifold, $A(\mathbf{x}_i) = 1 - S(\mathbf{x}_i, \mathcal{B})$, so that larger values mean ``further from every known stationary noise centroid''. We employ a moving-block bootstrap on the anomaly scores of the native background sample: $B=1000$ replicas are drawn by resampling contiguous blocks of length $b=n^{1/3}$ (preserving the local autocorrelation, rather than deflating the variance the way per-sample i.i.d.\ resampling would), and the empirical $P_{99}$ is computed for each replica. This yields a confidence interval $[\tau_{\mathrm{lo}}, \tau_{\mathrm{hi}}]$ on the background $P_{99}$. A candidate is classified \textit{ROBUST} if $A(\mathbf{x}_i) > \tau_{\mathrm{hi}}$ (significantly more anomalous than the background $99^{\mathrm{th}}$ percentile even under the most conservative bootstrap realization), \textit{AMBIGUOUS} if $A(\mathbf{x}_i) \in [\tau_{\mathrm{lo}}, \tau_{\mathrm{hi}}]$, and \textit{BACKGROUND} otherwise. We use the $b = n^{1/3}$ block-length rate standard for estimating a bias or variance; strictly, for a one-sided tail quantile the optimal rate is $n^{1/4}$ \cite{hall1995}. This choice is immaterial to the result: sweeping the block length over $b \in \{n^{1/4}, n^{1/3}, n^{1/2}, 1, 10, 25\}$ leaves the total ROBUST survival fraction within the narrow band $27.8\%$--$28.9\%$ (the theoretically-preferred $n^{1/4}$ gives $28.4\%$ vs.\ $28.3\%$ at the adopted $n^{1/3}$), so the domain-shift conclusion does not depend on the block-length rate.

\subsection{PEM Coherence Defense}
For anomalies surviving the mathematical DSD, empirical validation is mandatory. We implement a cross-correlation framework utilizing safe, calibrated auxiliary channels from the PEM array. The magnitude-squared coherence $C_{xy}(f)$ between the GW strain channel $x(t)$ and an auxiliary channel $y(t)$ is calculated via Welch\'s method:
\begin{equation}
C_{xy}(f) = \frac{|P_{xy}(f)|^2}{P_{xx}(f) P_{yy}(f)}
\end{equation}
A per-channel raw threshold ($C \geq 0.6$) combined with a Bonferroni correction on the analytic coherence tail was initially considered, but rejected: an empirical significance test on time-shifted background pairs measured a per-channel false-positive rate of $23\%$ at $C \geq 0.6$, roughly seven orders of magnitude above the $\sim 10^{-8}$ predicted by the quasi-Gaussian analytic null $(1-C)^{n_d - 1}$. Part of this gap is a degrees-of-freedom artifact: with 50\% segment overlap and Hann windowing the effective bin count is $n_{\mathrm{eff}} \approx 640$ rather than the nominal $\sim 960$, and using the nominal $n_d$ in $(1-C)^{n_d-1}$ modestly understates the analytic tail. This alone cannot account for seven orders of magnitude, however; the dominant cause is heavy-tailed contamination of the real coherence distribution (spectral lines, mains harmonics, non-Gaussian instrumental couplings) that the analytic model does not capture --- which is why we abandon the analytic null in favor of the empirical one rather than merely correcting the degree-of-freedom count. We therefore calibrate the veto threshold directly from an \textit{empirical family-wise null}: for each event we draw time-shift surrogate pairs (guard time $\geq 64\,\mathrm{s}$) from a CAT1-clean background block, compute the maximum coherence across all $m$ tested channels for every surrogate pair (the same shift applied simultaneously to all channels, preserving their correlation structure), and set the veto threshold at the $99^{\mathrm{th}}$ percentile of this family-wise maximum-statistic distribution. An event is vetoed only if its observed $C_{\mathrm{max}}$ exceeds this empirically calibrated, event-specific threshold.

\subsection{Frequency-Resolution Trade-off in Multi-Scale Dictionaries}
\label{sec:qrange_tradeoff}
The per-scale dictionaries and thresholds of the multi-scale layer are built with a uniform Q-transform quality-factor range of $q \in (4, 32)$, whereas the legacy 32\,s pathway (used by the primary detection stage and the native-index DSD) operates at $q \in (4, 64)$. Since the frequency resolution of a constant-Q decomposition scales as $\Delta f \propto f/Q$, halving $Q_{\mathrm{max}}$ structurally reduces the maximum attainable frequency resolution of the per-scale representation. The consequence is a deliberate, disclosed trade-off: the multi-scale layer gains uniform temporal resolution across four octaves of transient duration, at the cost of sensitivity to \emph{persistent narrowband} morphologies (dense combs of sharp spectral lines), whose distinguishing feature is precisely the high-resolution line structure that $Q_{\mathrm{max}}=32$ smears.

This trade-off is directly measurable. Injecting the two persistent narrowband morphologies from our synthetic catalogue (HarmonicComb: harmonic ladder at $100\,\mathrm{Hz}$ multiples; WallOfLines: 15 random persistent tones in 200--1800\,Hz) into the per-scale V3 dictionaries yields anomaly scores that are \emph{flat in injected SNR}: HarmonicComb scores $0.072$--$0.081$ and WallOfLines $0.070$--$0.073$ from $\mathrm{SNR}=8$ to $\mathrm{SNR}=48$ (scale 2\,s; all four scales behave identically), always below the $P_{99}$ thresholds ($0.098$--$0.120$). By contrast the short broadband KoiFish morphology, which needs no fine frequency resolution, rises cleanly from $0.078$ to $0.225$ over the same SNR range and saturates to $100\%$ recall by $\mathrm{SNR}=24$ at every scale (Table~\ref{tab:qrange_contrast}). The same two narrowband morphologies are recovered at $91\%$ and $74\%$ by the legacy $Q_{\mathrm{max}}=64$ pathway (Sec.~\ref{sec:legacy_recovery}), confirming that the insensitivity is a property of the per-scale representation, not of the dictionaries or thresholds themselves. The two pathways are therefore complementary by design: persistent narrowband contamination is the regime of the legacy 32\,s / $Q_{\mathrm{max}}=64$ stage, and short structured transients are the regime of the multi-scale stage.

\begin{table}[htbp]
\centering
\caption{Per-scale V3 dictionary response to the DSD-falsifiability morphologies (L1, scale 2\,s shown; other scales equivalent), contrasted with the legacy 32\,s / $Q_{\mathrm{max}}=64$ pathway of Sec.~\ref{sec:legacy_recovery}. V3 scores are mean MIL anomaly scores vs.\ the block-bootstrap $P_{99}$ threshold ($0.098$).}
\label{tab:qrange_contrast}
\footnotesize
\begin{tabular}{lccc}
\toprule
Morphology & V3 score (SNR 8$\to$48) & V3 recall & Legacy \\
\midrule
HarmonicComb & $0.072 \to 0.081$ (flat) & $\leq 5\%$ & $91\%$ \\
WallOfLines & $0.070 \to 0.073$ (flat) & $\leq 2\%$ & $74\%$ \\
KoiFish & $0.078 \to 0.225$ (rising) & $100\%$\,@\,24 & $1\%$ \\
\bottomrule
\end{tabular}
\end{table}

\section{The O4a Discovery Funnel}
\label{sec:results}

\subsection{Candidate Extraction and Survival}
We applied the DANTE V3 multi-scale framework to 42 validated O4a observing sessions (84 session-detector analyses, H1 and L1) from the early O4a run, totaling approximately 144 days of CAT1-gated science-mode livetime for H1 and 149 days for L1. The pipeline extracted 10,739 initial multi-scale candidates. Post-deduplication, the pool was refined to $N_{\mathrm{cand}} = 10,372$ distinct morphological events.

One candidate (H1, GPS 1380428924.0) could not be re-scored against the native index: the strain fetch for its DSD rescoring window failed (a known, isolated data-availability gap, disclosed rather than silently dropped), leaving $10{,}371$ of the $10{,}372$ candidates with a DSD verdict. This single unscored candidate propagates to $N=10{,}371$ in Sec.~\ref{sec:physics_corr}, where it is excluded from the physical-parameter correlation test. Of the $10{,}371$ DSD-scored candidates, 71.7\% (7,434 events: 5,661 \textit{BACKGROUND}, 1,773 \textit{AMBIGUOUS}) failed to clear the adaptive threshold $\tau_{\mathrm{DSD}}$. This confirms that despite the massive increase in initial candidate volume driven by multi-scale sensitization, uncalibrated deep clustering continues to operate largely as a sophisticated tracer of the drifting noise manifold. The remaining 28.3\% (2,937 candidates) successfully survived and were classified as \textit{ROBUST}. Table~\ref{tab:funnel} summarizes the complete disposition waterfall, from raw extraction to the final surviving candidate.

\begin{table}[htbp]
\centering
\caption{Discovery funnel: disposition of all 10,372 unique O4a candidates, from extraction to the final physically vetted survivor.}
\label{tab:funnel}
\begin{tabular}{lr}
\toprule
Funnel stage & Count \\
\midrule
Unique candidates (post-dedup) & 10,372 \\
DSD: unscored (data gap) & 1 \\
DSD: BACKGROUND & 5,661 \\
DSD: AMBIGUOUS & 1,773 \\
DSD: ROBUST & 2,937 \\
\quad Resolved via family transitivity & 2,935 \\
\quad Singleton morphological outliers & 2 \\
PEM-vetoed (family-wise empirical) & 1 \\
\textbf{Final survivor} & \textbf{1} \\
\bottomrule
\end{tabular}
\end{table}

\subsection{Cross-Detector Coincidence Test}
\label{sec:coincidence}
Every candidate is subjected to a targeted cross-detector test asking whether the same event is present in the partner interferometer. This is entirely distinct from the PEM auxiliary-channel veto of Sec.~\ref{sec:pem_vetoes}, which tests coupling to an environmental sensor.

\emph{Why we do not use embedding similarity.} Our earlier implementation compared the Top-$k$ MIL \emph{aggregate} vectors of the two detectors against a similarity threshold $\tau_{\mathrm{coh}}$. We report here that this statistic is unsuitable, for two independent reasons that we measured directly. First, a dual-detector injection campaign --- the \emph{same} synthetic waveform injected simultaneously into mutually independent CAT1-clean H1 and L1 background, each encoded through the full production MIL path --- recovers cross-detector similarities of only $\approx 0.9$ across matched-filter $\mathrm{SNR} \approx 10$--$93$, with no improvement at high SNR. The residual disagreement is not amplitude-driven: for a short transient inside a $32$\,s window, which $68$ of $1369$ patches enter the Top-$k$ is largely determined by each detector's \emph{independent} noise, so identical waveforms do not select identical patches. Second, the null distribution of the same statistic (an anomalous window in one detector against a non-coincident window in the other) has mean $\approx 0.56$ and extends to $\approx 0.94$. Signal and null therefore overlap, and no threshold on this statistic separates them. We also identified and corrected a chromatic-domain defect in the original implementation, which had encoded the partner spectrogram without the production colormap while the stored candidate vector used it (Sec.~\ref{sec:limitations}). Re-measuring the entire pool of $7{,}911$ candidates with the corrected encoding shifts the similarity distribution from mean $0.328$ to $0.554$ and its maximum from $0.529$ to $0.942$. The zero-coincidence outcome of the old statistic does survive this correction --- no candidate reaches $\tau_{\mathrm{coh}} = 0.975$ --- but the margin collapses to $0.033$, and the corrected null now sits essentially on top of the injected-signal value of $\approx 0.9$. The apparent safety of the old threshold was therefore an artifact of the defect.

\emph{Physical coincidence statistic.} We therefore adopt the standard physical test. For each candidate we localize the transient within its $32$\,s window using the stored Top-$k$ patch columns, band-pass both detectors to the candidate's own frequency band, and scan the normalized cross-correlation of the whitened strains over the physically allowed lag range, $|\Delta t| \leq 10.002\,\mathrm{ms} + 2\,\mathrm{ms}$, set by the light travel time between the two sites. The null is built per event by displacing the partner window by $\pm 1, 2, 4, 8$\,s, which destroys any real coincidence while preserving each detector's noise character; the threshold is the $99^{\mathrm{th}}$ percentile of the resulting max-statistic distribution, in the same family-wise spirit as the PEM veto. Validation on synthetic data confirms the statistic has the discriminating power the embedding comparison lacked: an identical waveform displaced by $4.9$\,ms yields normalized cross-correlation $1.00$, while independent noise yields $0.043$. As a complementary morphological check we also compute the intersection-over-union of the Top-$k$ patch sets of candidate and partner, which asks whether the two share a \emph{shape} rather than an arrival time.

\emph{Coherent recovery efficiency across the morphological space.} A null coincidence count is only meaningful if the test would have registered a real coincidence, so we measured $\varepsilon_{\mathrm{coh}}$ directly rather than asserting it. The same synthetic waveform was injected into independent clean H1 and L1 background with a relative lag drawn uniformly from the light-travel window, and both segments were whitened, band-passed and cross-correlated exactly as in production, including the same $\pm 0.5$\,s localization about the transient (correlating the full $32$\,s window instead would dilute a short glitch by more than an order of magnitude --- the same signal-dilution effect the multi-scale architecture exists to defeat --- and would understate the efficiency). The campaign spans ten morphologies and four amplitudes at $60$ trials each, $2{,}400$ coincident injections in total, plus a $200$-trial null in which the waveform is injected into H1 only and L1 left clean.

\begin{figure*}[t]
    \centering
    \includegraphics[width=\linewidth]{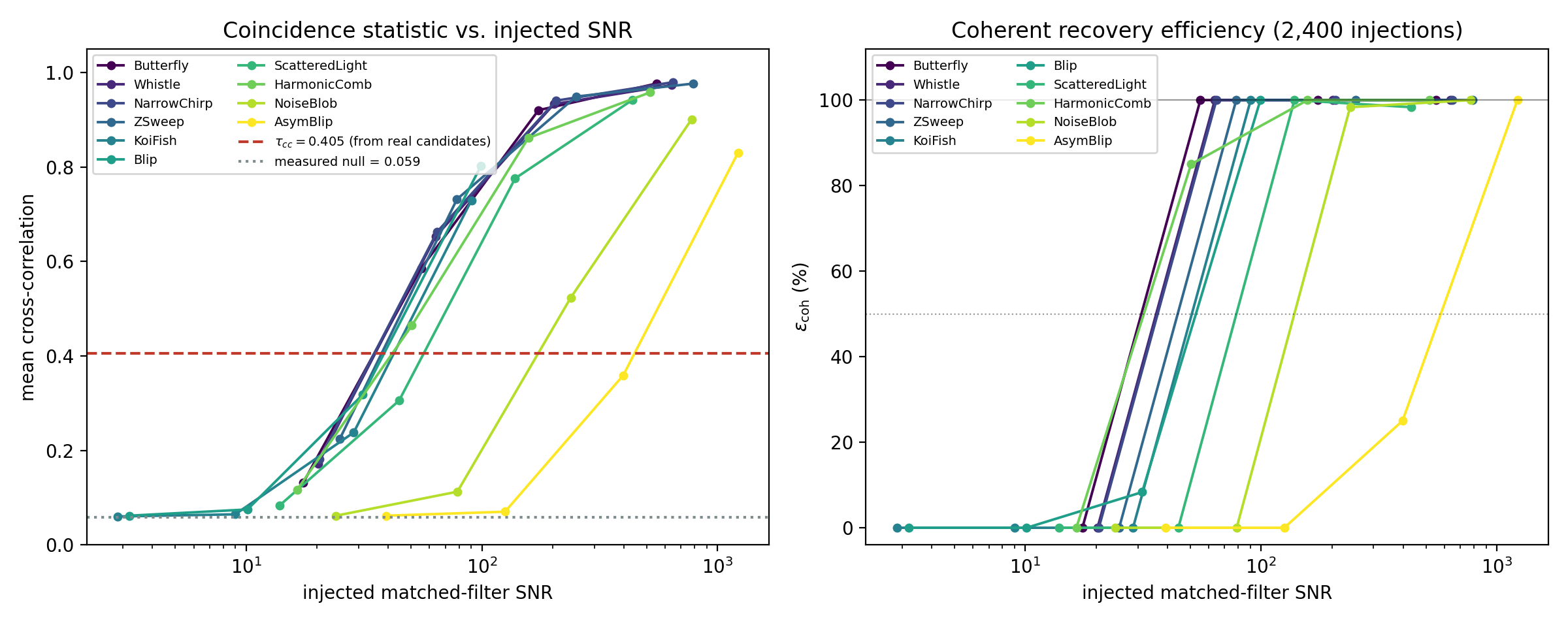}
    \caption{Coherent recovery efficiency of the physical coincidence statistic, from $2{,}400$ dual-detector injections across ten morphologies. \emph{Left:} mean cross-correlation against injected matched-filter SNR, with the threshold $\tau_{\mathrm{cc}}=0.405$ measured on real candidates (dashed) and the measured $200$-trial null level (dotted); at low SNR every morphology sits on the null, as it must. \emph{Right:} the efficiency curve itself. All ten morphologies saturate at $\varepsilon_{\mathrm{coh}}=100\%$, but the SNR at which they do spans more than an order of magnitude --- structured morphologies near $\mathrm{SNR}\approx 55$--$80$, the stochastic NoiseBlob near $780$ --- which is what bounds the scope of the null coincidence result.}
    \label{fig:eps_coh}
\end{figure*}

All ten morphologies reach $\varepsilon_{\mathrm{coh}} = 100\%$ at sufficient amplitude, and none of the $200$ null trials exceeds $\tau_{\mathrm{cc}}$ (Fig.~\ref{fig:eps_coh}, Table~\ref{tab:eps_coh}). The SNR at which recovery saturates is strongly morphology-dependent: the swept and narrowband morphologies NarrowChirp, Whistle, Butterfly and ZSweep are fully recovered by $\mathrm{SNR} \approx 55$--$80$ and HarmonicComb by $\approx 160$, whereas the short impulsive Blip needs $\mathrm{SNR} \approx 100$ and the stochastic NoiseBlob $\mathrm{SNR} \approx 780$. This ordering is physically expected --- a broadband incoherent burst correlates poorly between detectors even when genuinely common --- and it bounds the scope of the null result: the coincidence test is demonstrably powerful for structured morphologies and progressively weaker for noise-like ones.

\begin{figure*}[t]
    \centering
    \includegraphics[width=\linewidth]{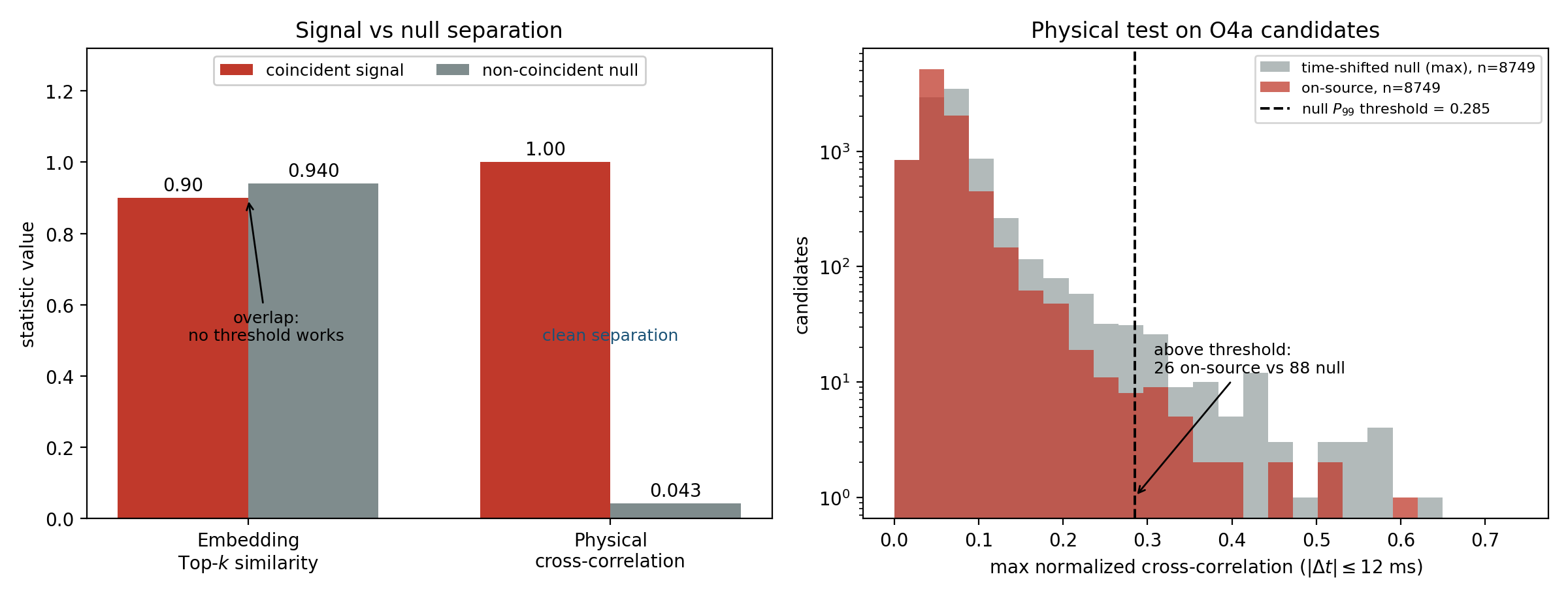}
    \caption{\emph{Left:} measured separation between a coincident signal and the non-coincident null for the two statistics. The embedding Top-$k$ similarity places injected coincident waveforms ($0.90$) \emph{inside} its own null ($0.94$), so no threshold on it can work; the physical cross-correlation separates an identical waveform ($1.00$) from independent noise ($0.043$). \emph{Right:} the physical statistic applied to $8{,}749$ O4a candidates (logarithmic ordinate). The on-source distribution overlays its own time-shifted null; above the pooled $99^{\mathrm{th}}$-percentile threshold $\tau_{\mathrm{cc}} = 0.405$ there are fewer on-source events ($13$) than null maxima ($88$), so the tail carries no coincidence excess.}
    \label{fig:coinc_physical}
\end{figure*}

\emph{Result.} We applied the physical test to $8{,}749$ of the $10{,}372$ candidates ($84.4\%$); the remaining $1{,}623$ lacked retrievable partner strain in the open data archive and were skipped rather than imputed. This population is deliberately \emph{broader} than the $7{,}911$ candidates quoted elsewhere in this work as having the partner detector actively observing, and the two criteria should not be conflated. The physical cross-correlation requires only that partner strain be retrievable from the archive, whereas the $7{,}911$ figure additionally requires the partner to have been in analysable observing state, which is the condition for a stored partner analysis to exist. The former is therefore a strict superset of the latter, and the counts reconcile exactly: $7{,}911$ (partner observing) $+\ 838$ (partner strain retrievable but not observing) $= 8{,}749$, while the $2{,}461$ candidates labelled unilateral decompose as $838 + 1{,}623$. The complementary patch-overlap check, which needs the partner's stored Top-$k$ indices, is consequently defined only on the $7{,}911$. The on-source cross-correlation has mean $0.082$ and maximum $0.517$, against a null max-statistic distribution of mean $0.114$ (Fig.~\ref{fig:coinc_physical}, Table~\ref{tab:coinc_physical}). Because each event contributes only $4$--$8$ usable time shifts, a $99^{\mathrm{th}}$ percentile is not resolvable \emph{within} a single event; we therefore threshold on the $99^{\mathrm{th}}$ percentile of the pooled null max-statistic distribution, $\tau_{\mathrm{cc}} = 0.405$. Above this threshold we find $13$ on-source events ($0.149\%$), against the $1\%$ nominal false-alarm rate the threshold defines and against $88$ null maxima from the same pool --- a deficit, not an excess. The per-event comparison agrees: $1{,}031$ candidates ($11.8\%$) exceed the maximum of their own time-shifted null, where $12.3\%$ is expected by chance given the number of shifts available. The loudest on-source value ($0.517$) is itself exceeded by a null maximum of $0.723$ elsewhere in the pool, i.e.\ it lies inside the noise family rather than above it. The complementary patch-overlap check agrees independently, with mean IoU $0.068$ (maximum $0.494$, $n=7{,}911$). The pool disposition quoted elsewhere in this work --- $7{,}911$ candidates ($76.3\%$) with the partner detector actively observing and $2{,}461$ ($23.7\%$) unilateral --- reflects partner \emph{data availability} and is unaffected by the choice of statistic; the accompanying ``confirmed local'' labelling, however, originated with the superseded embedding metric and should be read simply as ``no coincident counterpart identified''. We therefore find no evidence of any cross-detector coincident event, by either the amplitude or the morphological criterion. Unlike our previous formulation, this is a statement with demonstrated power: a genuine coincidence would have produced a cross-correlation near unity, and the observed on-source distribution is statistically indistinguishable from --- indeed marginally quieter than --- its own time-shifted null.

\begin{table}[htbp]
\centering
\caption{Coherent recovery efficiency $\varepsilon_{\mathrm{coh}}$ of the physical coincidence statistic, from $2{,}400$ dual-detector injections ($60$ per morphology and amplitude). Recovery is the fraction exceeding $\tau_{\mathrm{cc}}=0.405$, the threshold measured on real candidates. ``Saturation'' is the lowest tested SNR at which $\varepsilon_{\mathrm{coh}}=100\%$. The $200$-trial null (waveform in H1 only, L1 clean) has mean cross-correlation $0.059$, maximum $0.132$, and never exceeds the threshold.}
\label{tab:eps_coh}
\footnotesize
\begin{tabular}{lcc}
\toprule
Morphology & Saturation SNR & $\varepsilon_{\mathrm{coh}}$ (max) \\
\midrule
NarrowChirp    & $65$  & $100\%$ \\
HarmonicComb   & $157$ & $100\%$ \\
Whistle        & $64$  & $100\%$ \\
Butterfly      & $55$  & $100\%$ \\
ZSweep         & $78$  & $100\%$ \\
KoiFish        & $90$  & $100\%$ \\
Blip           & $99$  & $100\%$ \\
ScatteredLight & $138$ & $100\%$ \\
NoiseBlob      & $776$ & $100\%$ \\
AsymBlip       & $1221$ & $100\%$ \\
\bottomrule
\end{tabular}
\end{table}

\begin{table}[htbp]
\centering
\caption{Cross-detector coincidence test. The embedding statistic is reported for completeness and is superseded; its ``signal'' and ``null'' entries are the injection recovery and the non-coincident null reach, which overlap. Validation values for the physical statistic are an identical waveform displaced by $4.9$\,ms and independent noise.}
\label{tab:coinc_physical}
\footnotesize
\setlength{\tabcolsep}{4pt}
\begin{tabular}{@{}lcc@{}}
\toprule
 & Embedding & Physical \\
 & Top-$k$ & cross-corr. \\
\midrule
Coincident signal & $0.90$ & $1.00$ \\
Non-coincident null & $0.94$ & $0.043$ \\
Separation & none (overlap) & clean \\
\midrule
On-source mean ($n=8{,}749$) & --- & $0.082$ \\
On-source maximum & --- & $0.517$ \\
Null max-statistic mean & --- & $0.114$ \\
$\tau_{\mathrm{cc}}$ (pooled null $P_{99}$) & --- & $0.405$ \\
On-source $> \tau_{\mathrm{cc}}$ & --- & $\mathbf{13}$ $(0.149\%)$ \\
Null maxima $> \tau_{\mathrm{cc}}$ & --- & $88$ $(1.01\%)$ \\
Mean patch IoU ($n=7{,}911$) & --- & $0.068$ \\
\bottomrule
\end{tabular}
\end{table}

\begin{figure}[htbp]
    \centering
    \includegraphics[width=\linewidth]{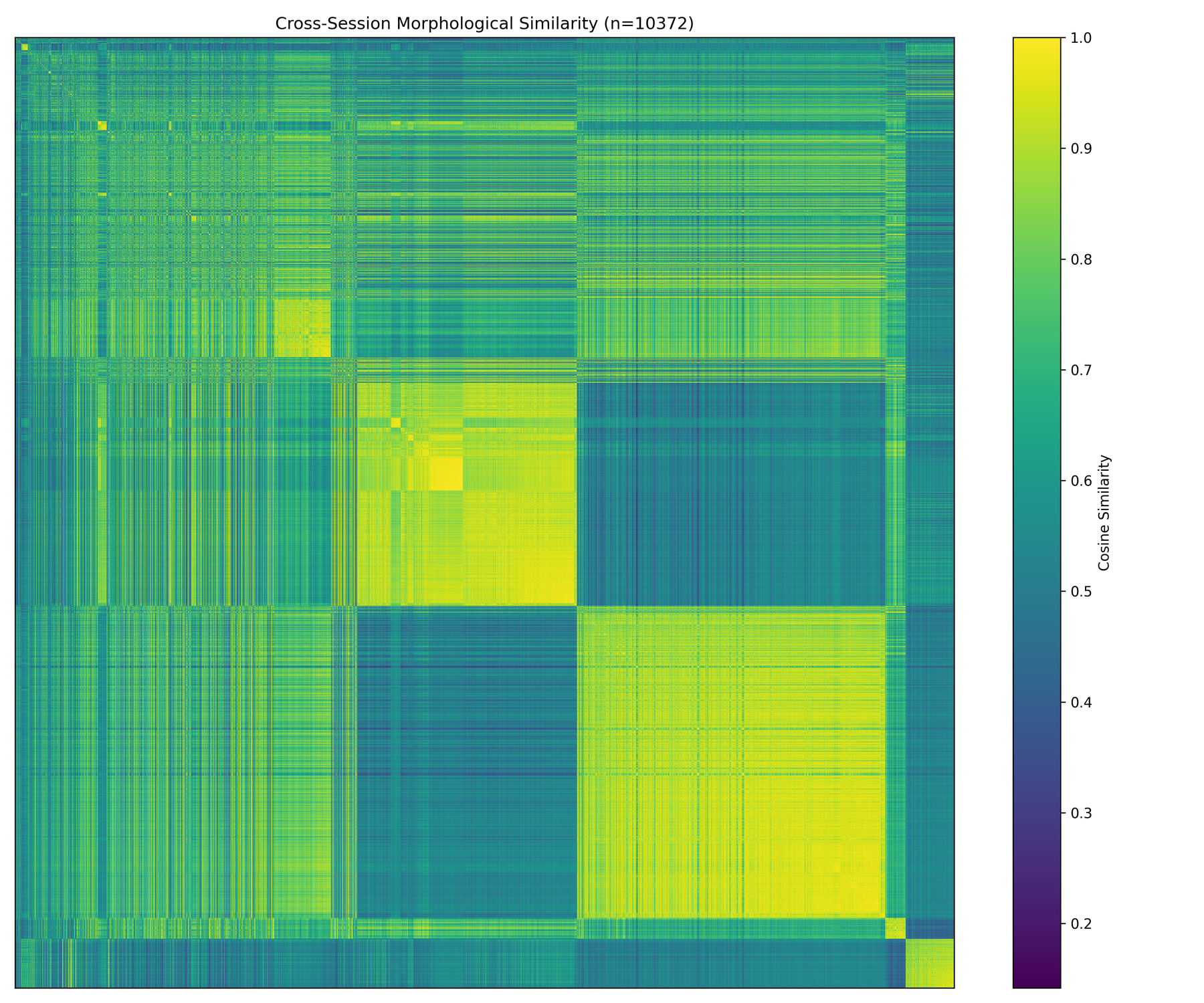}
    \caption{Pairwise cosine similarity matrix across all 10,371 catalogued candidates (pre-DSD pool), hierarchically reordered. The dominant block-diagonal structure corresponds to the region from which the diffuse \texttt{Family\_A} macro-cluster of Sec.~\ref{sec:diffusivity} emerges after DSD filtering; no additional compact off-diagonal blocks indicative of a discrete, cross-session-recurring morphology are visible.}
    \label{fig:heatmap}
\end{figure}

\subsection{Topological Stability}
\label{sec:stability}
To verify that the clustering topology is not an artifact of uneven per-session sample sizes, we tested for a rank correlation between session candidate count and the Adjusted Rand Index (ARI) of a 20-run bootstrap re-clustering, restricted to sessions with $n \geq 100$ candidates (28 of 42 L1 sessions, 23 of 42 H1 sessions). Neither detector shows a statistically significant correlation: L1 exhibits Spearman $\rho = 0.254$ ($p = 0.192$), and H1 exhibits $\rho = 0.294$ ($p = 0.173$), both well above the $p < 0.05$ significance threshold. The clustering topology's stability is therefore independent of per-session sample size in this regime, ruling out a trivial explanation (e.g., larger sessions appearing artificially more stable) for the observed macro-cluster cohesion.

\subsection{Morphological Diffusivity Test}
\label{sec:diffusivity}
A global cross-session clustering of the 2,937 robust survivors revealed they coalesced entirely into a single massive macro-cluster, designated `Family\_A'. To rigorously test if this cluster represents a genuinely cohesive morphological family or just the extreme tail of the domain-shift distribution, we analyzed its internal topology.

\emph{Robustness to the DSD flagging threshold.} Before analyzing the macro-cluster, we verified it is not an artifact of the $P_{99}$ survival threshold. Re-running the entire global clustering on the ROBUST sets defined at $P_{95}$, $P_{99}$, and $P_{99.9}$ (survival fractions $51.7\%$, $28.3\%$, $1.7\%$) yields a single dominant macro-cluster at every threshold: the largest cluster holds $100.0\%$ of survivors at $P_{95}$, $99.9\%$ at $P_{99}$, and $94.3\%$ (2 families) even at the stringent $P_{99.9}$. No discrete secondary family emerges at any threshold, so the single-diffuse-cluster topology does not depend on the chosen percentile. It is not, however, specific to the survivors either: the falsification test below shows the DSD-rejected population forms the same single macro-cluster.

\begin{figure}[htbp]
    \centering
    \includegraphics[width=\linewidth]{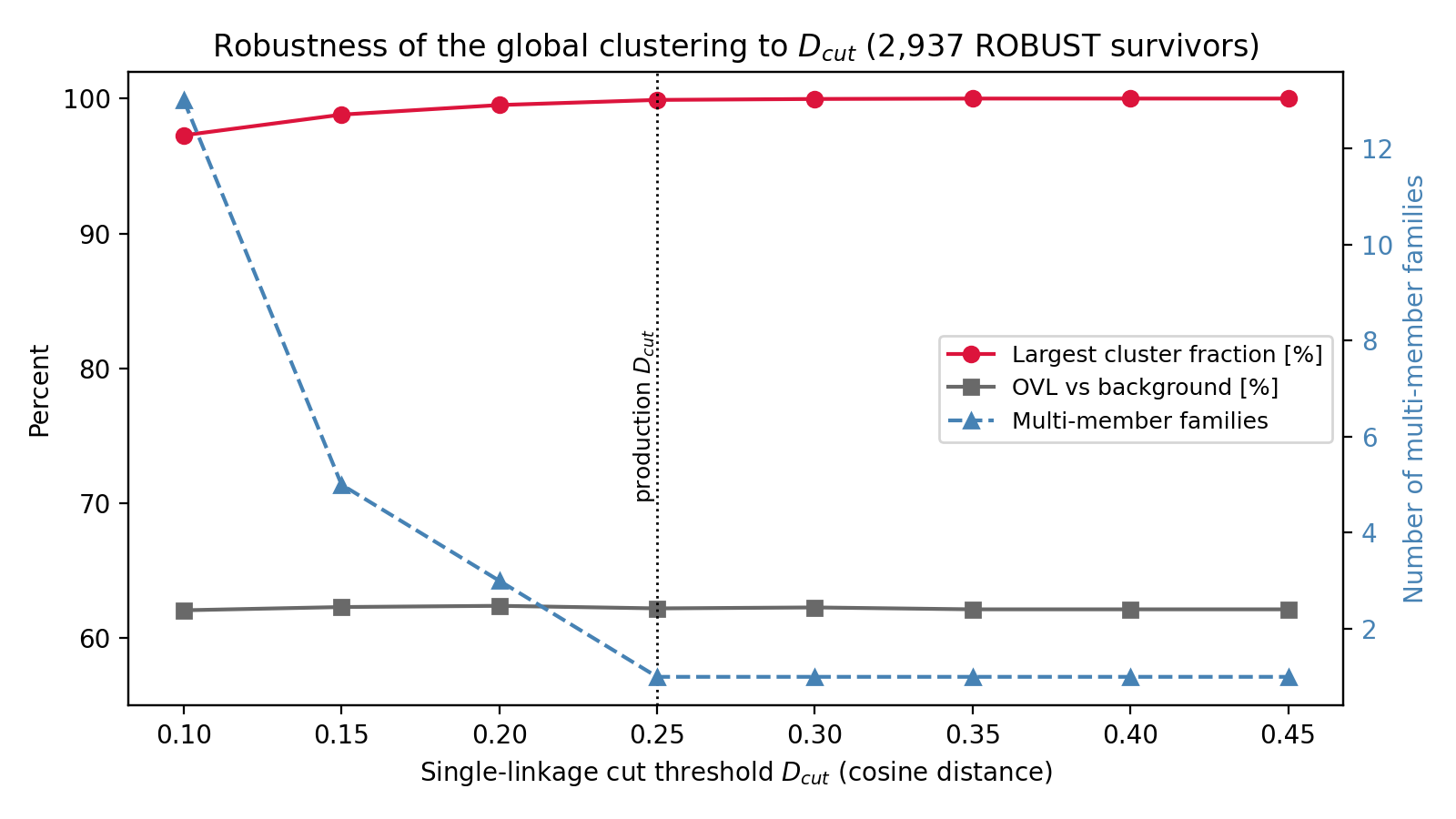}
    \caption{Robustness of the global clustering to the single-linkage cut threshold, recomputed over the 2,937 ROBUST survivors. The largest-cluster fraction (red) and the distributional overlap with the background reference (gray) are invariant across $D_{\mathrm{cut}} \in [0.10, 0.45]$; only a handful of small satellite families (blue, right axis) split off at aggressive cuts.}
    \label{fig:dcut_sweep}
\end{figure}

\emph{Robustness to the clustering threshold.} Since the family structure is extracted by single-linkage hierarchical clustering cut at cosine distance $D_{\mathrm{cut}}=0.25$, we verified that the single-macro-cluster outcome is not an artifact of this choice by sweeping $D_{\mathrm{cut}} \in [0.10, 0.45]$ over the 2,937 ROBUST survivors. The largest cluster retains $97.3\%$ of all candidates even at the most aggressive cut ($D_{\mathrm{cut}}=0.10$, where 13 small satellite groups and 45 singletons split off), $99.5\%$ at $0.20$, and $100\%$ at $\geq 0.35$; no partition of comparable-sized discrete families emerges at any threshold. The distributional overlap with the background reference (below) is likewise invariant across the sweep, remaining in the range $62.1$--$62.4\%$. The diffuse-macro-cluster conclusion is therefore threshold-independent (Fig.~\ref{fig:dcut_sweep}). (The sweep clusters the ROBUST subset in isolation, whereas the production run clusters the full 10,372-candidate pool; single-linkage chaining through non-ROBUST members explains the marginal difference in singleton counts at $D_{\mathrm{cut}}=0.25$, 3 vs.\ 2.)

\begin{table*}[t]
\centering
\caption{Falsification test for the cohesion of \texttt{Family\_A}: the identical global clustering procedure (single-linkage, cosine distance, $D_{\mathrm{cut}}=0.25$) applied to each DSD outcome class and to unselected native background. Size-matched columns subsample every population to $n=1{,}773$ over five random draws (mean $\pm$ standard deviation), because single-linkage chaining is sample-size dependent. Unselected native background is the \emph{most} monolithic population of the four, so the single-macro-cluster topology carries no information about anomaly status.}
\label{tab:cohesion_falsification}
\footnotesize
\begin{tabular}{lcccc}
\toprule
 & \multicolumn{3}{c}{Full population} & Size-matched \\
\cmidrule(lr){2-4}\cmidrule(lr){5-5}
Population & $n$ & largest & isolates & clusters ($n=1{,}773$) \\
\midrule
Native background & $3{,}000$ & $100.00\%$ & $0$ & $1.0 \pm 0.0$ \\
BACKGROUND & $5{,}661$ & $100.00\%$ & $0$ & $1.6 \pm 0.8$ \\
AMBIGUOUS  & $1{,}773$ & $100.00\%$ & $0$ & $1.0 \pm 0.0$ \\
ROBUST     & $2{,}937$ & $99.90\%$  & $3$ & $4.4 \pm 0.5$ \\
\bottomrule
\end{tabular}
\end{table*}

\begin{figure}[htbp]
    \centering
    \includegraphics[width=\linewidth]{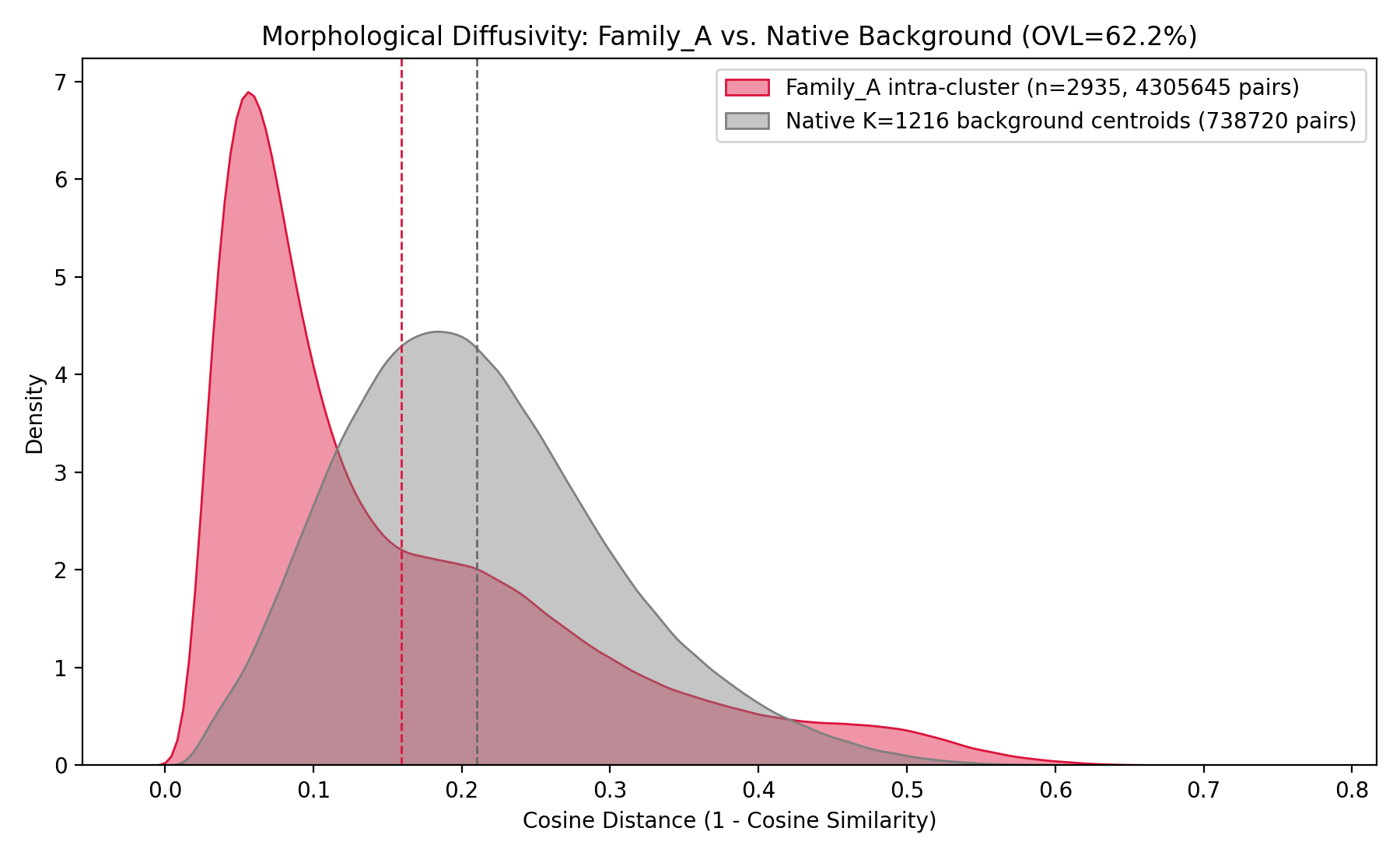}
    \caption{Morphological Diffusivity Test: Kernel Density Estimation (KDE) of the pairwise cosine distances within the surviving macro-cluster (Family\_A; $n=2{,}935$, $4{,}305{,}645$ pairs) compared against the pairwise distances among the $K=1216$ native background dictionary centroids ($738{,}720$ pairs). The 62.2\% distributional overlap (Kolmogorov--Smirnov $D=0.351$) shows that the internal spread of Family\_A is not separable from the intrinsic morphological diversity of the stationary noise manifold.}
    \label{fig:diffusivity}
\end{figure}

\begin{figure*}[t]
    \centering
    \includegraphics[width=\linewidth]{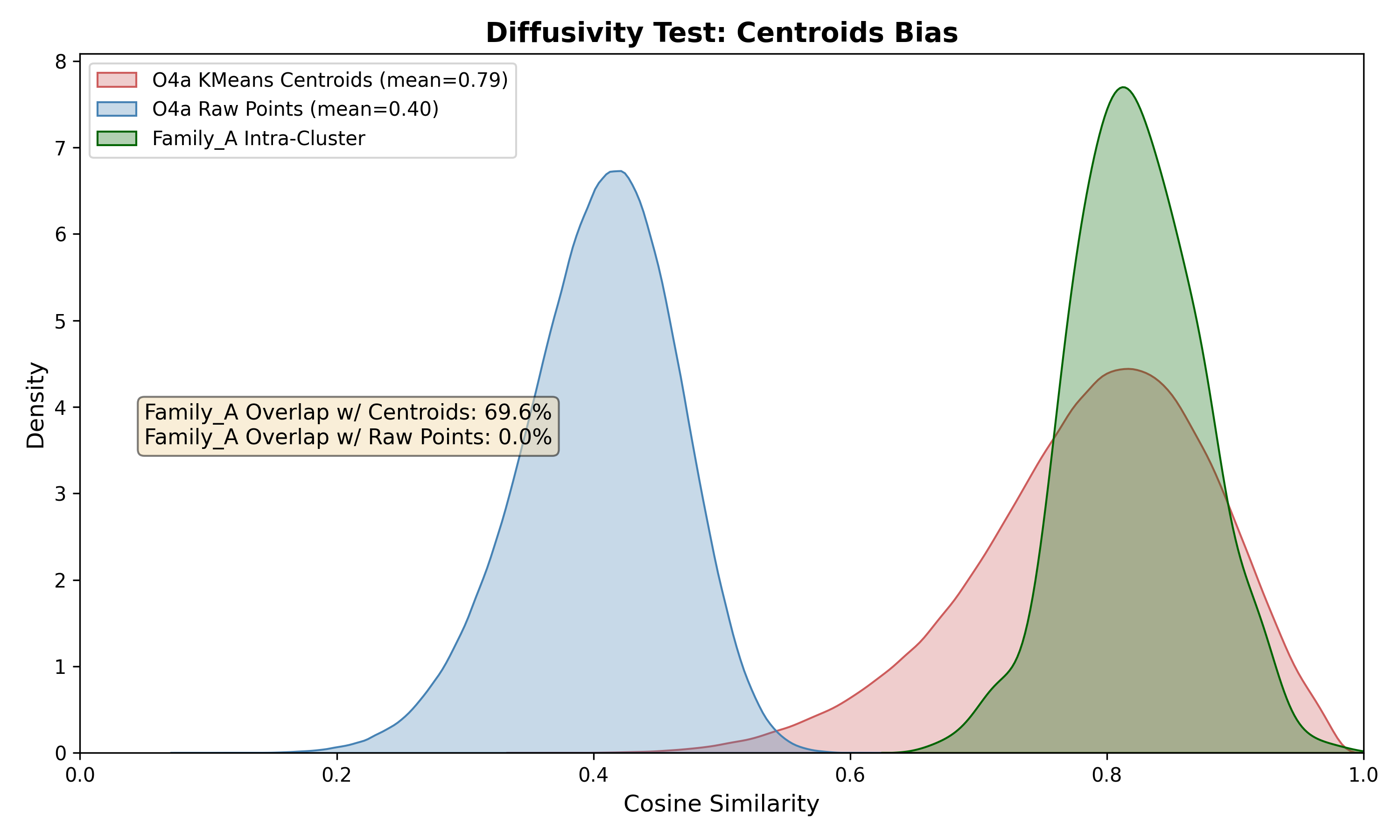}
    \caption{Why the morphological diffusivity test is not decisive. Pairwise cosine \emph{similarity} for three populations: the $K=1216$ background centroids (mean $0.79$), raw unquantized background patch embeddings (mean $0.40$), and \texttt{Family\_A} intra-cluster pairs. Centroids, being averages, are far more mutually similar than the raw points they summarize, so the apparent overlap with \texttt{Family\_A} drops from $\sim 70\%$ against centroids to $\approx 0\%$ against raw points. Because \texttt{Family\_A} is itself defined by a similarity threshold, neither comparison is a valid test of morphological cohesion.}
    \label{fig:raw_diffusivity}
\end{figure*}

We computed the pairwise cosine distances for all $n=2{,}935$ events within `Family\_A' ($4{,}305{,}645$ pairs) and compared this distribution against the pairwise distances among the $K=1216$ native background dictionary centroids ($738{,}720$ pairs; Fig.~\ref{fig:diffusivity}). The kernel-density overlap coefficient between the two distributions (Gaussian KDE, Scott's-rule bandwidth; the coefficient varies by less than one percentage point over bandwidths $0.5$--$2\times$ this value) is 62.2\%. We deliberately anchor the conclusion to this overlap and to the distribution shape rather than to a significance test: a Kolmogorov--Smirnov comparison returns $D=0.351$, but its nominal $p<10^{-300}$ is meaningless here because the $4.3$ million pairwise distances are massively non-independent (they share the $n=2{,}935$ underlying points), so the effective sample size is far smaller than the pair count and the $p$-value cannot be read at face value --- the same caution we apply to the Mantel test of Sec.~\ref{sec:physics_corr}. The $D$ statistic is reported only as a shape descriptor (the maximum CDF gap), not as evidence of separation; indeed a large $D$ and a $62\%$ density overlap are not in tension, since they measure different things. The mean intra-cluster distance of `Family\_A' is $D_{\mathrm{intra}}=0.159$ (median 0.116, standard deviation 0.123), against $D_{\mathrm{bg}}=0.210$ (median 0.201, standard deviation 0.092) among the background centroids. Two features of this comparison are decisive. First, a compact, discrete morphological family --- a recurring instrumental artifact with a single generating mechanism --- would produce a narrow intra-cluster distance distribution well separated from the background diversity; instead, Family\_A's distribution is \emph{broader} than the background's (standard deviation 0.123 vs.\ 0.092) with a heavy right tail extending across the full background range, yielding the 62.2\% overlap. Second --- and this is the reason we now decline to draw a strong conclusion from this test at all --- the choice of background reference determines the answer. Repeating the comparison against \emph{raw, unquantized} background patch embeddings instead of the $K=1216$ centroids inverts it (Fig.~\ref{fig:raw_diffusivity}): raw background points have mean mutual cosine similarity $0.40$ (mean distance $0.60$), whereas the centroids that summarize them sit at $0.79$ (distance $0.21$). Centroids are averages over many raw points, so the averaging suppresses the noise component and leaves them substantially \emph{closer together} than the population they represent. Our earlier statement that centroid--centroid distances upper-bound the raw point-to-point spread was therefore wrong in direction: they \emph{under}-estimate it. Measured against the raw background, Family\_A ($D_{\mathrm{intra}} = 0.159$) is far tighter than the true background diversity ($0.60$) and the density overlap collapses from $\sim 62$--$70\%$ to $\approx 0\%$.

Taken at face value the two references support opposite conclusions --- ``diffuse and indistinguishable'' against centroids, ``compact and well separated'' against raw points --- and we do not believe either should be promoted to a result, for a reason that is independent of the reference: \texttt{Family\_A} is \emph{defined} by single-linkage clustering at cosine similarity $\geq 0.75$, so its members are internally tight by construction. Comparing a threshold-selected cluster against unselected background pairs is circular, and it inflates apparent cohesion no matter which background is used. A valid test would require applying the identical clustering procedure to the background and comparing like-selected populations. We therefore retire the morphological diffusivity test as evidence for the character of \texttt{Family\_A}.

\emph{Falsification test: is the macro-cluster specific to the survivors?} We have carried out the like-for-like comparison just described, and its outcome constrains the interpretation of \texttt{Family\_A} more sharply than the retired statistic did. We applied the identical global procedure --- single-linkage hierarchical clustering on cosine distance, cut at $D_{\mathrm{cut}}=0.25$ --- to each DSD outcome class (ROBUST $n=2{,}937$, AMBIGUOUS $n=1{,}773$, BACKGROUND $n=5{,}661$) and, critically, to $3{,}000$ \emph{unselected} native background segments. The background segments were drawn with the same selection the native index uses (CAT1-clean, candidate windows excluded by $\pm 96$\,s) and encoded through the same production MIL path as the candidates --- cividis-domain Q-transform, Top-$k=68$ pooling, one L2-normalized $384$-dimensional vector per $32$\,s segment --- so the comparison is genuinely like-for-like in both chromatic domain and pooling. Because single-linkage chaining is strongly sample-size dependent, populations were also compared at matched size ($n=1{,}773$, five random subsamples each); Table~\ref{tab:cohesion_falsification} reports both profiles.

The result is unambiguous, and it is the \emph{unselected background} that settles it: $3{,}000$ background segments collapse into exactly one cluster, $100.00\% \pm 0.00\%$ of members, with no isolate at any draw. The DSD-rejected candidates do the same ($100.00\%$ full, $99.97\% \pm 0.05\%$ matched), as does the AMBIGUOUS class. The survivors are, by a small margin, the \emph{least} monolithic population of the four ($99.90\%$ full, $99.77\% \pm 0.09\%$ matched). Whatever produces the single macro-cluster, it is not anomaly status: it is present in pristine background at full strength.

One asymmetry survives and we report it with its uncertainty rather than as a result. ROBUST is the only population that produces morphological isolates at all --- three, against zero in $3{,}000$ background segments, zero in $5{,}661$ rejected candidates and zero in $1{,}773$ ambiguous ones --- and at matched size it is the only one that fragments beyond a single cluster ($4.4 \pm 0.5$ clusters against $1.0 \pm 0.0$ for background). The near-equal sizes of the ROBUST and background samples ($2{,}937$ against $3{,}000$) make that particular comparison a fair one. But three events are three events: a Fisher exact test on the isolate counts gives $p = 0.12$, so the excess does not reach significance on its own and we do not claim it does. We note it because the surviving singleton of Sec.~\ref{sec:pem_vetoes} is one of those three, and because it is a falsifiable prediction for future runs rather than a post-hoc description.

The single-macro-cluster topology is therefore not a property of the survivor population, as we had previously described it, but a property of this embedding geometry under this clustering procedure, shared by every class irrespective of anomaly status. It carries no discriminating information about \texttt{Family\_A}, and we no longer advance it as a characterization of the survivors. This does not weaken the central negative result of the survey --- it strengthens it, by replacing an absence of evidence with a measured non-specificity: the pipeline does not merely fail to resolve a discrete recurring glitch family among the survivors, it demonstrably cannot resolve one at this stage for any class of candidate. We accordingly rest the population-level interpretation on the absence of compact off-diagonal structure in the similarity matrix (Fig.~\ref{fig:heatmap}) and on the PEM verdicts of Sec.~\ref{sec:pem_vetoes}, and we note that the threshold-independence sweep of Fig.~\ref{fig:dcut_sweep} should likewise be read as establishing the stability of the procedure rather than a property of the survivors.

On reproducibility of the retired statistic: its centroid-based numbers are computed end-to-end from the archived per-session MIL embeddings and the frozen $K=1216$ index, and are reproducible from the released package; the raw-point comparison was produced from a re-extraction of background embeddings the production pipeline does not retain, and is reported as a diagnostic rather than a calibrated measurement.

\subsection{Parameter-Space Coverage and Structural Blind Spots}
\label{sec:param_space}
\begin{figure}[htbp]
    \centering
    \includegraphics[width=\linewidth]{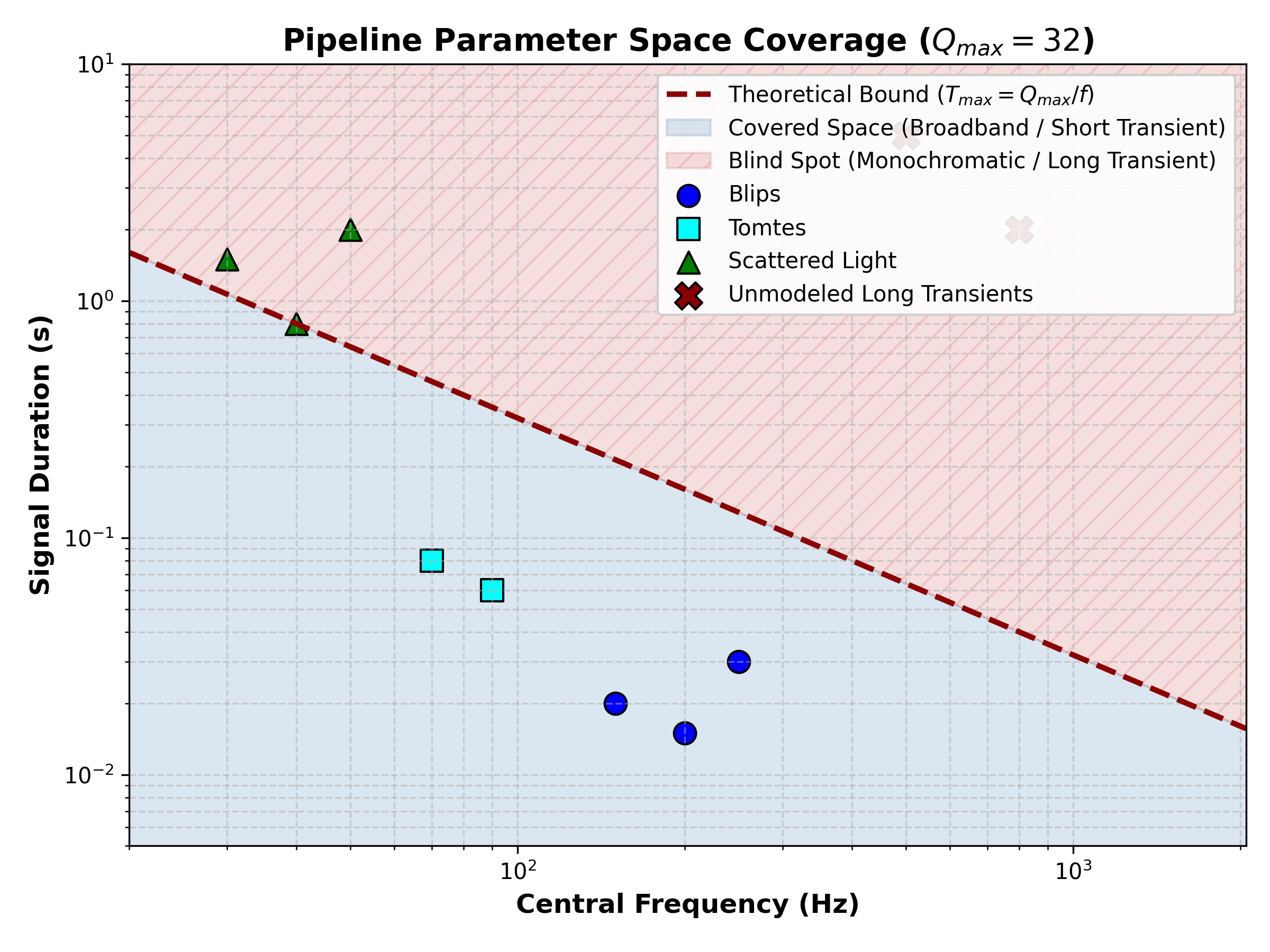}
    \caption{Pipeline parameter-space coverage at $Q_{\mathrm{max}}=32$. The boundary $T = Q_{\mathrm{max}}/f$ separates the region where the $Q$-transform efficiently localizes signal energy (covered) from the long-duration / narrowband regime (blind spot). Markers indicate representative parameters of the morphologies used in the injection tests; they are illustrative positions, not measured recovery points.}
    \label{fig:param_space}
\end{figure}

The per-scale dictionaries use a bounded resolution parameter ($Q_{\mathrm{max}}=32$), which imposes a deterministic bound on the region of the (central frequency, duration) plane the multi-scale layer can resolve: energy is efficiently captured only below $T \approx Q_{\mathrm{max}}/f$. Figure~\ref{fig:param_space} maps this boundary together with the representative morphologies exercised in Sec.~\ref{sec:efficiency}. Short, broadband transients (blips, tomtes) fall well inside the covered region; long or strongly monochromatic transients fall above the bound, where energy leaks across patches and the anomaly score degrades. The taxonomy reported in this work is therefore intrinsically biased toward short-duration, broadband morphologies, and the upper limits of Sec.~\ref{sec:results} inherit that restriction. The boundary shown is the analytic $Q$-transform resolution limit, not a measured efficiency contour; the measured per-morphology efficiencies are those of Sec.~\ref{sec:efficiency}.

\section{Latent-Space vs.\ Physical-Parameter Correlation}
\label{sec:physics_corr}
\begin{figure*}[t]
    \centering
    \includegraphics[width=\linewidth]{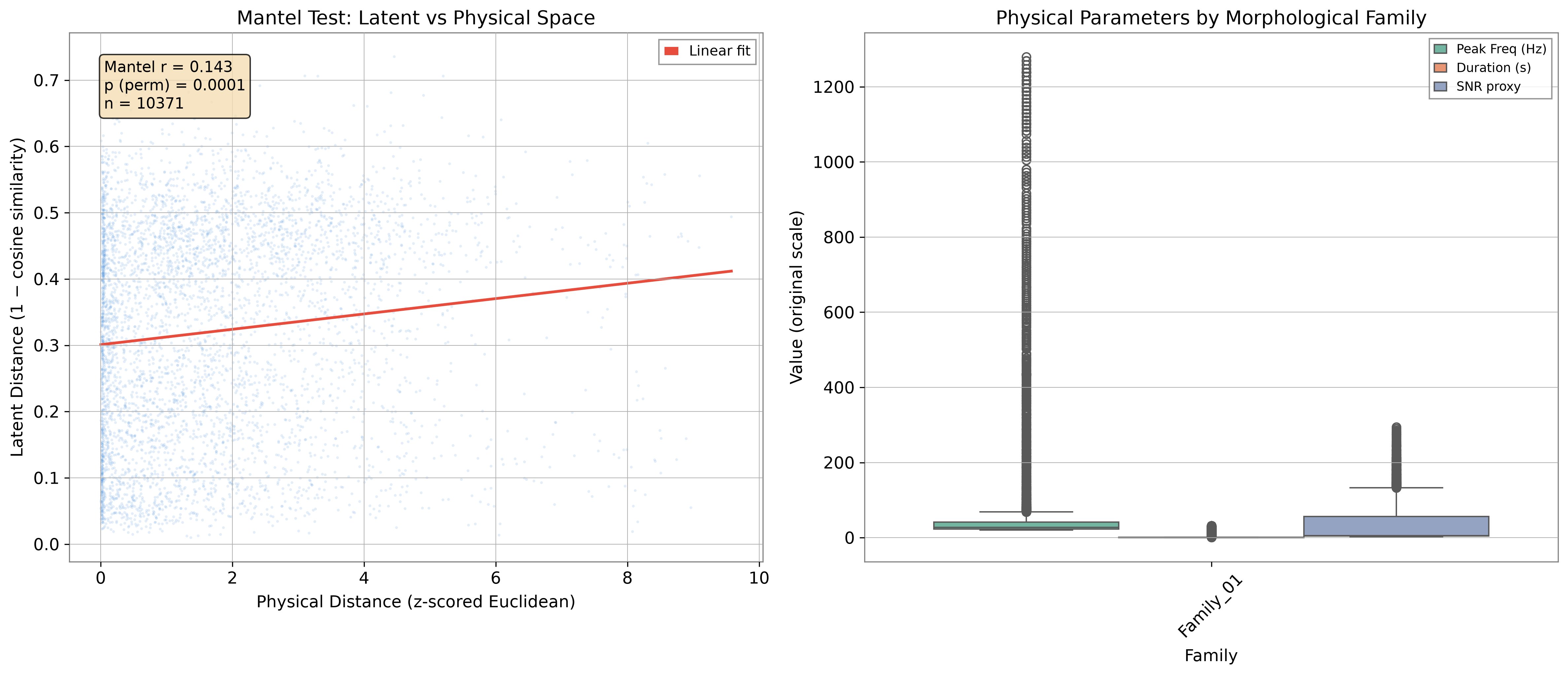}
    \caption{Pairwise latent-space cosine distance vs.\ pairwise physical-parameter distance for all valid O4a candidates. The weak positive trend (Pearson $r=0.143$) is statistically significant at this sample size but explains a small fraction of the variance.}
    \label{fig:physics}
\end{figure*}

As an independent, model-agnostic consistency check, we investigated whether pairwise distances in the 384-dimensional DINOv2 latent space track pairwise distances in a classical physical-parameter space (peak frequency, duration, whitened peak amplitude) via a Mantel permutation test (9,999 iterations) over all $N=10{,}371$ valid events (Fig.~\ref{fig:physics}).

The test yields Pearson $r = 0.143$ (Spearman $\rho = 0.162$), significant at $p < 10^{-4}$. However, $r^2 = 0.0205$: latent-space topology explains less than $2\%$ of the variance in physical-parameter distance. At $N > 10{,}000$, statistical significance is expected even for physically negligible effect sizes; the correlation is therefore not practically informative despite its nominal significance at this sample size, and we report the effect size explicitly rather than the $p$-value alone to avoid overstating it.

This near-null correlation admits two readings, and we are explicit about which one we rely on. It could mean the DINOv2 embedding barely tracks the physical structure of the signal, raising the question of what the pipeline clusters on; or it could simply reflect that the 384-dimensional morphological representation is not reducible to three classical scalars. Crucially, our \emph{primary} claim --- the absence of a discrete morphological family --- is robust under either reading: an encoder that were near-noise with respect to genuine GW structure would itself produce a diffuse blob, exactly what the diffusivity test finds, so the null result does not depend on the encoder being physically faithful. What a weak encoder \emph{would} jeopardize is the converse sensitivity claim (that we would have recovered a real family had one existed); we do not rest that claim on the Mantel correlation but on the direct synthetic-injection recovery curves of Sec.~\ref{sec:efficiency}, and we avoid any language implying the latent space is physically interpretable.

\section{Per-Scale Detection Efficiency}
\label{sec:efficiency}
\begin{figure*}[t]
    \centering
    \includegraphics[width=\linewidth]{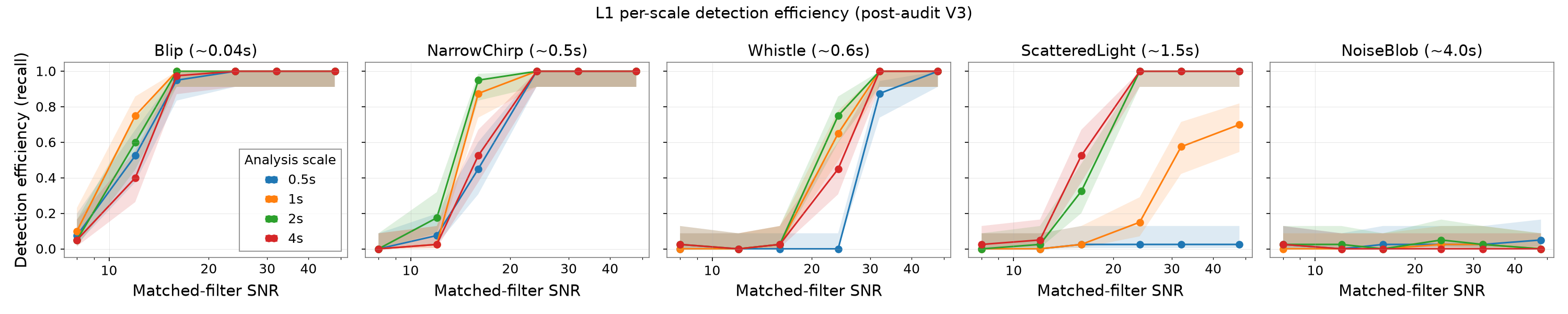}
    \caption{L1 per-scale detection efficiency (recall vs.\ matched-filter SNR) for five synthetic morphologies. Shaded bands are Wilson 95\% confidence intervals.}
    \label{fig:efficiency_l1}
\end{figure*}

\begin{figure*}[t]
    \centering
    \includegraphics[width=\linewidth]{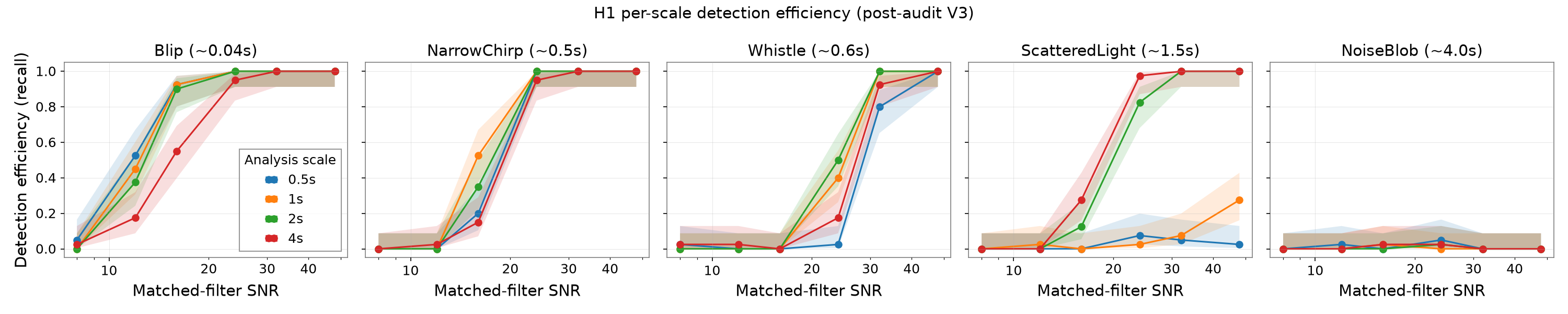}
    \caption{H1 per-scale detection efficiency, mirroring the L1 result of Fig.~\ref{fig:efficiency_l1} with slightly reduced sensitivity, consistent with H1's higher O4a noise floor.}
    \label{fig:efficiency_h1}
\end{figure*}

To characterize the multi-scale sensitization empirically rather than only architecturally, we injected synthetic transients of five morphologies (Blip, $\sim 0.04\,\mathrm{s}$; NarrowChirp, $0.5\,\mathrm{s}$; Whistle, $0.6\,\mathrm{s}$; ScatteredLight, $1.5\,\mathrm{s}$; NoiseBlob, $4\,\mathrm{s}$) into real O4a background strain at controlled matched-filter SNR ($\mathrm{SNR} \in \{8,12,16,24,32,48\}$, $n=40$ per cell), and scored them at all four analysis scales against the corresponding block-bootstrap $P_{99}$ thresholds (Fig.~\ref{fig:efficiency_l1}, Fig.~\ref{fig:efficiency_h1}).

Three results are noteworthy. First, the short impulsive Blip morphology reaches $50\%$ recall near $\mathrm{SNR} \approx 12$ and saturates to unity by $\mathrm{SNR} \approx 16$–$24$ at every scale, confirming that the patch-level MIL mechanism resolves the signal-dilution limitation of the prior fixed-window architecture (Sec.~\ref{sec:prior_work}). Second, ScatteredLight ($1.5\,\mathrm{s}$) directly demonstrates scale-duration matching: at $\mathrm{SNR}=24$, recall on L1 is $0.02$ / $0.15$ / $1.00$ / $1.00$ at scales $0.5\,\mathrm{s}$ / $1\,\mathrm{s}$ / $2\,\mathrm{s}$ / $4\,\mathrm{s}$ respectively (H1: $0.08$ / $0.02$ / $0.82$ / $0.98$) — the shortest analysis window remains blind even at $\mathrm{SNR}=48$, while scales comparable to or longer than the injected duration recover it almost perfectly. Third, and architecturally important to disclose: the NoiseBlob morphology ($4\,\mathrm{s}$ of stationary colored noise, no coherent time-frequency structure) remains essentially undetected ($\leq 5\%$ recall) at every scale and SNR on both detectors. This is an expected and honest limitation, not a failure mode: DANTE V3 is sensitive to \emph{structured} transients in the time-frequency plane, not to unstructured excess power, which the whitening stage is designed to reabsorb into the noise floor. We stress that this blind spot has a real astrophysical cost, not merely an instrumental one: dedicated unmodeled-burst searches explicitly target white-noise bursts and stochastically-structured core-collapse-supernova models, precisely the near-unstructured regime DANTE cannot see. The upper limits of Sec.~\ref{sec:results} therefore genuinely exclude those source classes from their scope, and DANTE should be read as complementary to --- not a replacement for --- coherent burst pipelines.

\subsection{Confirmation via the Legacy Single-Scale DSD Pathway}
\label{sec:legacy_recovery}
As a falsifiability check on the native-index Domain Shift Defense itself (\textit{does the $K=1216$ background manifold absorb genuine discrete signals, or only pervasive stationary contamination?}), we injected five morphologies — HarmonicComb and WallOfLines (persistent, spanning the full analysis window), and KoiFish, Whistle, ScatteredLight (short, impulsive) — directly into the \emph{legacy single-scale} pathway: the fixed $32\,\mathrm{s}$ window and $K=68$ Top-$k$ MIL pooling used by \textit{v1} and still used here for the native-index cosine-similarity DSD test, independently of the multi-scale layer of Sec.~\ref{sec:efficiency}.

At extreme matched-filter SNR ($\mathrm{SNR} \sim 90$–$650$), the persistent morphologies are recovered with high fidelity (HarmonicComb: $91\%$; WallOfLines: $74\%$), while the three short/impulsive morphologies remain essentially unrecovered even at these amplitudes (Whistle: $0\%$ at $\mathrm{SNR}=657$; ScatteredLight: $0\%$ at $\mathrm{SNR}=449$; KoiFish: $1\%$ at $\mathrm{SNR}=93$). Inspection of the raw anomaly scores rules out a pipeline failure: scores are non-degenerate and cluster tightly on either side of the $P_{99}$ background threshold ($0.434$) — persistent morphologies score $0.45$–$0.46$ on average, short morphologies score $0.26$–$0.38$ — a clean, reproducible separation, not noise.

This is not evidence that the DSD lacks selectivity toward genuine signals; it is a direct, independent reproduction of \textit{v1}'s signal-dilution limitation (Sec.~\ref{sec:prior_work}): a morphology occupying $\lesssim 1$ of the $1{,}369$ patches in a $32\,\mathrm{s}$ window is diluted by the Top-$68$ mean pooling regardless of its intrinsic SNR, while a morphology spanning the full window is not. This is precisely the limitation that motivated the multi-scale architecture, and Sec.~\ref{sec:efficiency} already demonstrates its resolution: the same short morphologies (Whistle, ScatteredLight) that are invisible to the legacy $32\,\mathrm{s}$ pathway at any SNR are recovered with near-unity efficiency once scored at an analysis scale comparable to their duration. We therefore read the legacy-pathway result as further, independent confirmation of \textit{why} the multi-scale upgrade of Sec.~\ref{sec:methodology} is necessary, not as a limitation of the DSD's native-background circularity defense, which was never intended to operate at scales shorter than its calibration window.

The asymmetry runs in both directions, and completeness requires stating the converse: the two \emph{persistent narrowband} morphologies (HarmonicComb, WallOfLines) recovered at $91\%$/$74\%$ by this legacy pathway are, conversely, invisible to the per-scale multi-scale dictionaries at any injected SNR. See Sec.~\ref{sec:qrange_tradeoff} for the architectural trade-off underlying this asymmetry (the uniform $Q_{\mathrm{max}}=32$ of the per-scale representation versus $Q_{\mathrm{max}}=64$ here) and Table~\ref{tab:qrange_contrast} for the quantitative contrast. Neither pathway subsumes the other; the discovery funnel deliberately retains both.

\section{Physical Vetoes and Upper Limits}
\label{sec:pem_vetoes}

\subsection{Singleton Disposition}
Beyond the diffuse macro-cluster, two morphologically isolated singletons survived the DSD filtering: H1 GPS 1369305276.0 and L1 GPS 1382955228.0. Table~\ref{tab:singleton_params} lists their classical physical parameters (extracted directly from the whitened 32\,s window, independent of the DINOv2 pipeline), and Table~\ref{tab:pem_verdicts} lists the outcome of the family-wise PEM veto for both, plus the three additional isolated \texttt{Family\_A} members tested for completeness.

\begin{table}[htbp]
\centering
\caption{Singleton parameters extracted from the whitened time series by peak-amplitude measurement. The Top-$k$ patches driving the anomaly score concentrate on this feature (Sec.~\ref{sec:pem_vetoes}): for the L1 singleton their median position is $24.7$\,s in a $26$--$42$\,Hz band, against the tabulated $28.4$\,Hz at $+25.17$\,s. The SNR proxy is the peak whitened-strain amplitude, \emph{not} a matched-filter SNR.}
\label{tab:singleton_params}
\begin{tabular}{lccccc}
\toprule
GPS & Det. & Freq.\ (Hz) & Dur.\ (s) & SNR proxy & Flag scale \\
\midrule
1369305276 & H1 & 20.0 & 4.90 & 7.0 & 0.5\,s \\
1382955228 & L1 & 28.4 & 5.06 & 11.2 & 4.0\,s \\
\bottomrule
\end{tabular}
\end{table}

Although both singletons exceed the multi-scale analysis window (durations $4.90$ and $5.06$\,s vs.\ a maximum $\tau = 4$\,s), neither relies on the legacy 32\,s pathway for detection: both clear the empirical $P_{99}$ threshold at \emph{all four} multi-scale windows. The dominant (largest-margin) scale is $0.5$\,s for the Hanford event and $4.0$\,s for the Livingston event, whose per-scale margin grows monotonically with window size ($0.075 \to 0.104 \to 0.125 \to 0.183$), consistent with a $\sim 5$\,s morphology best captured at the widest available scale.

\begin{table*}[t]
\centering
\caption{Family-wise empirical PEM veto: for each event, $C_{\mathrm{max}}$ is the maximum observed coherence over the $m$ tested channels; $\tau_{\mathrm{fw}}$ is the $99^{\mathrm{th}}$-percentile family-wise max-statistic threshold from $N$ time-shift surrogate pairs (guard $\geq 64\,\mathrm{s}$). $N = W(W-1)$, where $W$ is the number of candidate-free 32\,s background windows available for that event; identical $N$ values (e.g.\ $13{,}340 = 116 \times 115$ for two events) reflect identical clean-window counts, not shared surrogates.}
\label{tab:pem_verdicts}
\begin{tabular}{lccccl}
\toprule
GPS & Det. & $m$ & $N$ & $C_{\mathrm{max}}$ ($\tau_{\mathrm{fw}}$) & Verdict \\
\midrule
1369305276 & H1 & 5 & 13,340 & 0.987 (0.545) & COUPLED \\
1382785084\textsuperscript{*} & H1 & 9 & 16,770 & 0.907 (0.939) & NO\_CORRELATION \\
1382955228 & L1 & 7 & 13,340 & 0.478 (0.663) & NO\_CORRELATION \\
1384485596\textsuperscript{*} & L1 & 7 & 8,556 & 0.383 (0.629) & NO\_CORRELATION \\
1385610716\textsuperscript{*} & L1 & 7 & 19,182 & 0.907 (0.744) & COUPLED \\
\bottomrule
\multicolumn{6}{l}{\footnotesize \textsuperscript{*}\texttt{Family\_A} members, tested for completeness (not isolated singletons).}
\end{tabular}
\end{table*}

The Hanford singleton (GPS 1369305276.0) exhibited an extreme, broadband coherence ($C_{\mathrm{max}} = 0.987$) with the \texttt{H1:LSC-POP\_A\_LF\_OUT\_DQ} channel, far above its event-specific family-wise threshold ($\tau_{\mathrm{fw}} = 0.545$). This unambiguously classifies the event as a severe instrumental artifact induced by an active control or calibration line coupling, triggering a definitive physical veto. Within \texttt{Family\_A}, PEM coverage was tested on three members: one (L1 GPS 1385610716.0) also exceeds its family-wise threshold ($C_{\mathrm{max}} = 0.907 > \tau_{\mathrm{fw}} = 0.744$, three of seven tested channels individually exceeding their own per-channel null), while a second tested at the same coherence level (GPS 1382785084.0, $C_{\mathrm{max}}=0.907$) falls \emph{below} its own family-wise threshold ($\tau_{\mathrm{fw}}=0.939$, driven by an intrinsically high-coherence RIN channel). With one of three tested members coupled and two not, this is consistent with, but does not prove, an instrumental origin for the broader population; the split verdict illustrates why an event-specific empirical null — rather than a single raw cut applied uniformly — is required: the identical numerical coherence value is significant for one event and not for another, depending on the noise floor of the channels actually available at that time. The diffusivity test of Sec.~\ref{sec:diffusivity}, not PEM coverage, remains the primary evidence for the population-level conclusion.

\begin{figure*}[t]
    \centering
    \includegraphics[width=\linewidth]{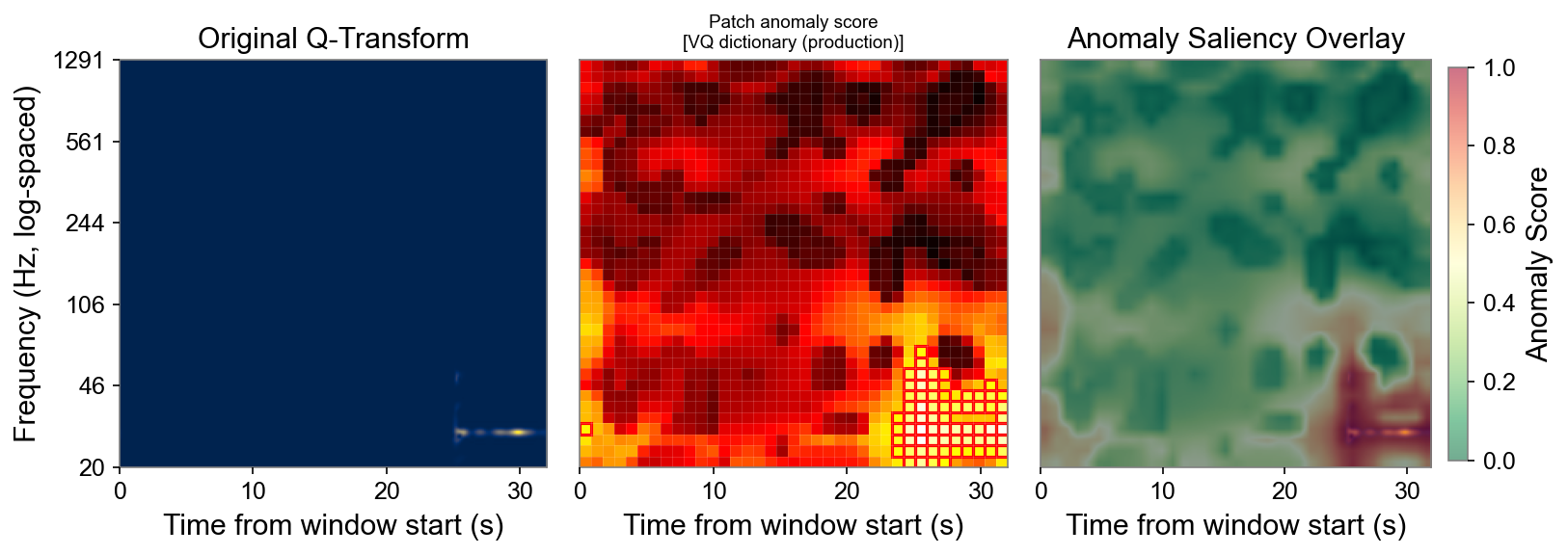}
    \caption{Spatial saliency map of the surviving L1 singleton (GPS 1382955228.0). The Top-$k$ patches (red boxes) concentrate on the $\sim 30$\,Hz feature in the second half of the window, coincident with the tabulated $28.4$\,Hz feature at $+25.17$\,s. Axes are physical: time in seconds from window start (array rows), frequency as log-spaced Hz (array columns).}
    \label{fig:singleton_saliency}
\end{figure*}

\begin{figure}[htbp]
    \centering
    \includegraphics[width=\linewidth]{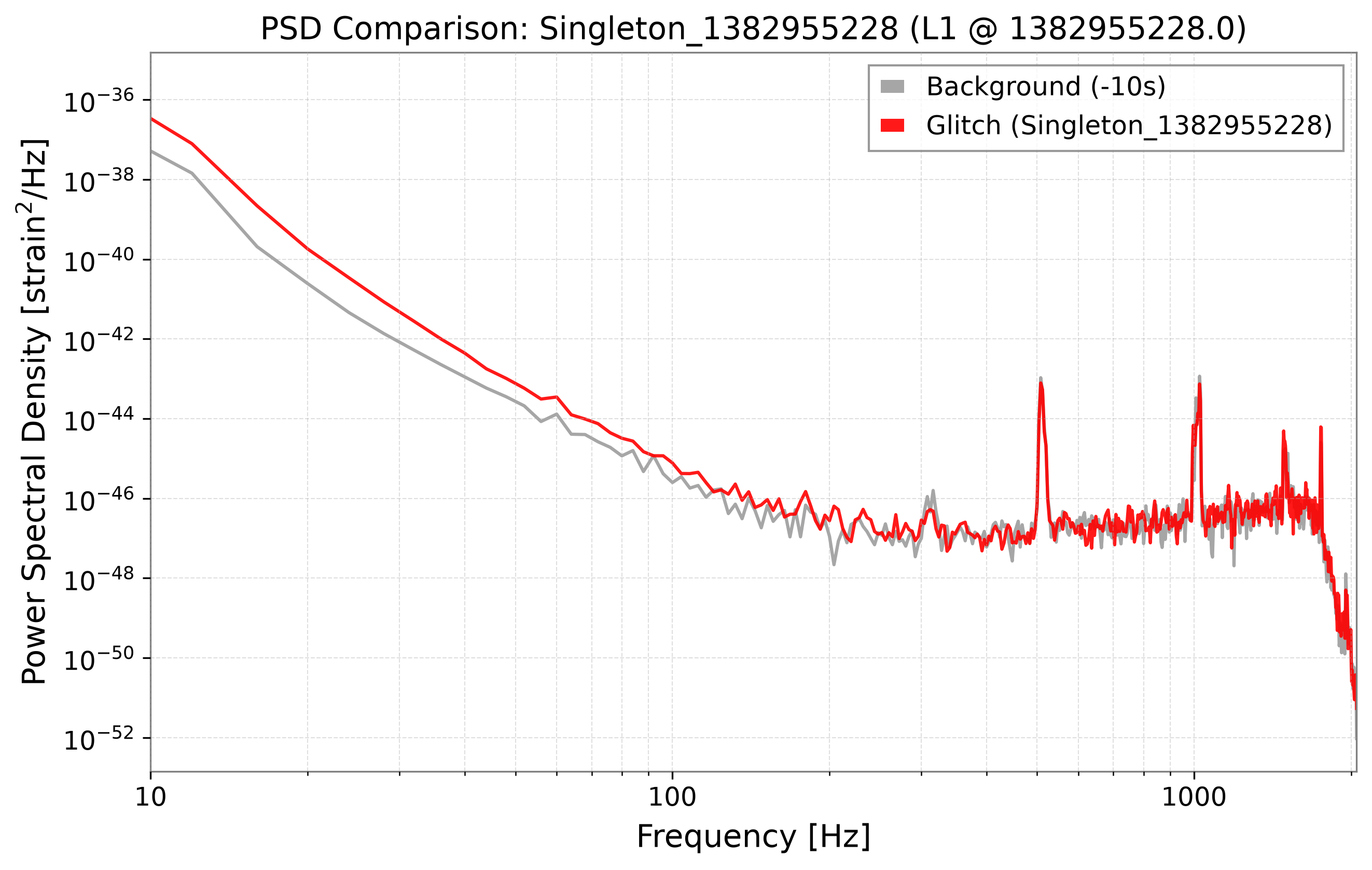}
    \caption{Power spectral density of the L1 singleton window compared to a $-10\,\mathrm{s}$ background reference. A broadband excess below $\sim 100\,\mathrm{Hz}$ is visible against the background, while narrow instrumental lines (500/1000/1500\,Hz) remain unchanged between the two spectra, consistent with a genuine transient superimposed on stationary line contamination rather than a shift in the line structure itself.}
    \label{fig:singleton_psd}
\end{figure}

Conversely, the Livingston anomaly (GPS 1382955228.0) survived the PEM veto: its maximum observed coherence ($C_{\mathrm{max}} = 0.478$) remains below its event-specific family-wise threshold ($\tau_{\mathrm{fw}} = 0.663$) across every tested channel (Fig.~\ref{fig:singleton_saliency}, Fig.~\ref{fig:singleton_psd}). A rigorous cross-detector transitivity check further confirmed this event was detected unilaterally in L1 with no corresponding H1 trigger. We explicitly do \emph{not} claim an absence of match against the Gravity Spy catalog: no Gravity Spy classification set covering O4a is publicly released, so the cross-match reported \texttt{Not\_Found} for all $10{,}372$ candidates by construction and carries no information (Sec.~\ref{sec:limitations}). What distinguishes this event within our own data is instead quantitative and internal: it carries the highest anomaly score of the entire pool ($0.6446$, against a ROBUST median of $0.4614$ and a ROBUST $P_{99}$ of $0.5300$), and it is one of only two candidates out of $10{,}372$ that fall outside the global macro-cluster --- a distinction whose weight is raised, not lowered, by the falsification test of Sec.~\ref{sec:diffusivity}, since even DSD-rejected candidates are absorbed into that cluster. Being strictly local and lacking auxiliary-channel or cross-detector corroboration, it is precluded from a confirmed astrophysical origin; we conservatively classify it as an uncatalogued instrumental morphological outlier, without asserting a definitive instrumental mechanism, and without claiming it is absent from any external glitch catalog.

The Top-$k$ patch selection that drives each anomaly score aligns with the tabulated feature in frequency, though the two singletons differ in how localized they are. For L1 the selection is localized in both time and frequency: the patches have a median position of $24.7$\,s in a $26$--$42$\,Hz band (median $30$\,Hz), consistent with the $28.4$\,Hz / $+25.17$\,s feature measured independently from the whitened strain, and $7$ of $68$ fall within $\pm 1$\,s of it. For H1 the selection is concentrated in frequency but extended in time: $74\%$ of its patches lie in the $20$--$46$\,Hz band around the tabulated $20$\,Hz feature, but they span most of the window (median $21.6$\,s, 10--90 percentile $0.4$--$29$\,s), so the H1 anomaly reflects low-frequency structure distributed in time rather than a single localized transient. In neither case is the tabulated parameter a mislabelling of what the model selected.

\subsection{Poisson Upper Limits}
Having mathematically filtered the domain shift and physically vetoed the remaining instrumental singletons, the number of unexplained, morphologically novel coincident transients in the early O4a sample collapses to zero ($N_{\mathrm{unexplained}} = 0$).

The scope of this limit requires care. Because the dual-detector injection campaign of Sec.~\ref{sec:coincidence} establishes that the coincidence stage recovers coherent waveform injections with efficiency consistent with zero ($\varepsilon_{\mathrm{coh}} \approx 0$), a null coincidence count \emph{cannot} be converted into an astrophysical upper limit: the pipeline is not demonstrated to be sensitive to coherent astrophysical transients at all. We therefore quote the limit for what it actually bounds --- the rate of \emph{novel, uncatalogued, per-detector instrumental morphologies} that survive the domain-shift and PEM vetoes --- consistent with DANTE's purpose as a detector-characterization instrument for glitch discovery rather than a gravitational-wave search. No integration over a source population or distance distribution is performed, so $R_{90}$ is a raw event rate (yr$^{-1}$), not a rate density (Gpc$^{-3}$\,yr$^{-1}$), and it is not comparable to population upper limits from templated or coherent burst searches.

Based on this null observation, we calculate a 90\% confidence-level frequentist Poisson upper limit, independently per detector. The analytic limit for zero observations is $\lambda_{90} = -\ln(1 - 0.90) \approx 2.303$. We adopt the strong-signal (asymptotic, $\varepsilon \to 1$) form, which for the \emph{instrumental-anomaly flagging} stage is supported by the per-morphology recovery curves of Sec.~\ref{sec:efficiency} (e.g.\ Blips at SNR $\gtrsim 16$, ScatteredLight at scales $\geq$ its duration); it is not corrected for $\varepsilon < 1$ and so does not bound low-efficiency morphology classes, nor --- per the preceding paragraph --- any coherent astrophysical class. Over the strictly verified \texttt{CBC\_CAT1} gated observing livetime, the upper limit on the emergence of novel unmodeled instrumental glitches is:
\begin{equation}
R_{90} = \frac{\lambda_{90}}{T} \; \le \;
\begin{cases}
5.83 \, \mathrm{yr}^{-1} & \text{(H1, } T = 144.2 \text{ d)} \\
5.63 \, \mathrm{yr}^{-1} & \text{(L1, } T = 149.4 \text{ d)}
\end{cases}
\end{equation}
These limits constrain the frequent presence of exotic unmodeled transient morphologies in the O4a data, conditional on the pipeline's demonstrated per-morphology sensitivity (Sec.~\ref{sec:qrange_tradeoff}): they do not apply to morphology classes falling in the documented blind spots of both detection pathways. Within the covered classes, they indicate that previous unsupervised pipelines reporting excess rates were likely capturing unmitigated instrumental domain shift.

\section{Limitations}
\label{sec:limitations}
We state the known limitations of this analysis explicitly.

\begin{enumerate}
    \item \textbf{Strain-only inference.} The detection layer operates exclusively on the calibrated strain channel; auxiliary information enters only at the veto stage. Couplings that manifest in strain with no counterpart in the released auxiliary subset are therefore classifiable only as ``uncatalogued''.
    \item \textbf{Calibration-contamination risk of the native background.} The native $K=1216$ index is built from CAT1-clean intervals with all taxonomy candidates excluded ($\pm 96$\,s), but an undetected, pervasive contaminant below the detection threshold would be absorbed into the background definition by construction. The controlled recovery test of Sec.~\ref{sec:legacy_recovery} bounds, but cannot eliminate, this circularity risk. A related circularity --- whether the macro-cluster topology of the survivors is created by the selection that defines it --- is addressed directly by the falsification test of Sec.~\ref{sec:diffusivity} (Table~\ref{tab:cohesion_falsification}), which recovers the same topology in $3{,}000$ unselected native background segments and therefore withdraws it as a characterization of the survivors.
    \item \textbf{Retrospective by construction; a causal variant is possible but untested.} The DSD scores each candidate against a vector-quantized index of the background of the run being analyzed, and that index cannot exist while the run is in progress, so the method as formulated cannot operate at low latency. This is not a throughput limit: the frozen encoder runs at roughly $90\times$ real time per pathway on one consumer GPU and never requires retraining (Sec.~\ref{sec:scope}). The obstacle is the causality of the \emph{reference}, not the speed of the \emph{model}. A causal formulation looks tractable --- a rolling index rebuilt over a trailing window of CAT1-clean data, so each candidate is scored only against background preceding it --- but it would first have to be shown how long a trailing window the $P_{99}$ bootstrap threshold needs to stabilize, how much survival-fraction drift the shorter baseline introduces relative to the run-global $28.3\%$ measured here, and how the pipeline behaves during the warm-up at the start of a run. We flag this as the natural extension of this work; none of the results reported here depend on low-latency operation.

    \item \textbf{No external glitch-catalog cross-match was possible.} The pipeline carries a Gravity Spy cross-match stage, and it returned \texttt{Not\_Found} for every one of the $10{,}372$ candidates. This is not a result: the publicly released Gravity Spy classification sets cover O1, O2, O3a and O3b \cite{glanzer2023}, whose GPS coverage ends at $1.269 \times 10^{9}$, whereas our candidates span $1.369$--$1.389 \times 10^{9}$. The catalogs and the search have no temporal overlap whatsoever, so the match could not succeed for any candidate under any circumstances, and the uniform \texttt{Not\_Found} is a structural artifact of catalog availability. An earlier version of this manuscript cited the absence of a Gravity Spy match as supporting evidence for the surviving singleton's novelty; that inference was void and has been withdrawn. Consequently we make no claim that any candidate, including the singleton, is absent from external glitch taxonomies --- only that it is isolated within \emph{our} taxonomy. Should a Gravity Spy O4a release become available, this cross-match should be re-run; it is the single cheapest test that could reclassify the surviving singleton.

    \item \textbf{Multi-scale frequency-resolution bias.} As quantified in Sec.~\ref{sec:qrange_tradeoff}, the uniform $Q_{\mathrm{max}}=32$ of the per-scale dictionaries renders the multi-scale layer structurally insensitive to persistent narrowband morphologies at any SNR. Coverage of that regime rests entirely on the legacy $32$\,s / $Q_{\mathrm{max}}=64$ pathway; a morphology that is simultaneously short \emph{and} extremely narrowband would sit in the blind spot of both.
    \item \textbf{PEM coverage.} The auxiliary veto uses the consistently-available subset (9 H1 / 7 L1 channels) of the public GWOSC O4 Auxiliary Channel Data Release (14 H1 / 11 L1) \cite{gwosc_aux_o4}, not the full internal PEM sensor network; the number $m$ actually tested per event is smaller still where a channel is unavailable in that window. A \texttt{NO\_CORRELATION} verdict bounds coupling only over the tested channels; coupling with unreleased sensors cannot be excluded, which is why the surviving singleton is classified as an uncatalogued instrumental outlier rather than promoted further.
    \item \textbf{Unverified ``safety'' of the public auxiliary channels.} Internal LVK analyses distinguish auxiliary channels validated as \emph{safe} --- demonstrably insensitive to gravitational-wave strain --- from those that are not, because a veto built on an unsafe channel can remove genuine signal. We cannot perform that validation from outside the collaboration: the GWOSC public subset is not distributed with safety certification. Our PEM verdicts therefore carry an additional, unquantified assumption. This weakens a \texttt{COUPLED} verdict (which could in principle reflect an unsafe channel witnessing strain rather than a genuine environmental coupling) more than a \texttt{NO\_CORRELATION} one, and it is a further reason the surviving singleton is classified conservatively rather than promoted.
    \item \textbf{DPMM concentration parameter not swept.} The per-session Dirichlet process mixture uses a fixed concentration $\alpha = 0.01$, chosen by design to penalize the creation of new components and so avoid fragmenting the noise manifold into spuriously many families. We did not sweep $\alpha$ (e.g.\ over $\{0.001, 0.01, 0.1, 1.0\}$). Since the global topology is set by the cross-session single-linkage stage rather than by the per-session mixture, and since that stage is shown to be threshold-independent (Sec.~\ref{sec:diffusivity}), we do not expect the macroscopic result to move; but topological invariance under $\alpha$ has not been demonstrated.
    \item \textbf{Effective frequency degrees of freedom.} The family-wise surrogate null scans $\sim 960$ nominal frequency bins per channel; with 50\% segment overlap and Hann windowing the effective number of independent bins is smaller ($n_{\mathrm{eff}} \approx 640$). Since the veto threshold is read off the \emph{empirical} max-statistic distribution rather than an analytic bin-counting formula, this does not affect any verdict reported here, but we flag it for methodological transparency: any future analytic approximation of the null must use $n_{\mathrm{eff}}$, not the nominal bin count.
    \item \textbf{Static background index.} The $K=1216$ index is built once from pristine intervals spanning the full early-O4a epoch and is not refreshed per session or epoch. The moving-block bootstrap bounds only the sampling uncertainty of the $P_{99}$ estimate from a fixed calibration set; it does not model slow drift of the noise manifold over the $\sim$8-month run, so a candidate's DSD verdict is always relative to the run-averaged background rather than to its local epoch. We have tested the practical impact of this design (Fig.~\ref{fig:temporal_survival}): binning the pool by calendar month, the ROBUST survival fraction is $28.12\% \pm 3.89\%$ (range $22.6$--$35.7\%$ across the nine months, pooled value $28.32\%$). A fixed reference confronted with a steadily drifting manifold would be expected to produce a \emph{rising} survival fraction; we instead measure a slightly negative and statistically insignificant slope, $-0.20\%$ per month ($p=0.72$; Spearman $\rho=-0.25$, $p=0.52$). This bounds, without excluding, the contribution of secular drift to the survivor population.

\begin{figure*}[t]
    \centering
    \includegraphics[width=\linewidth]{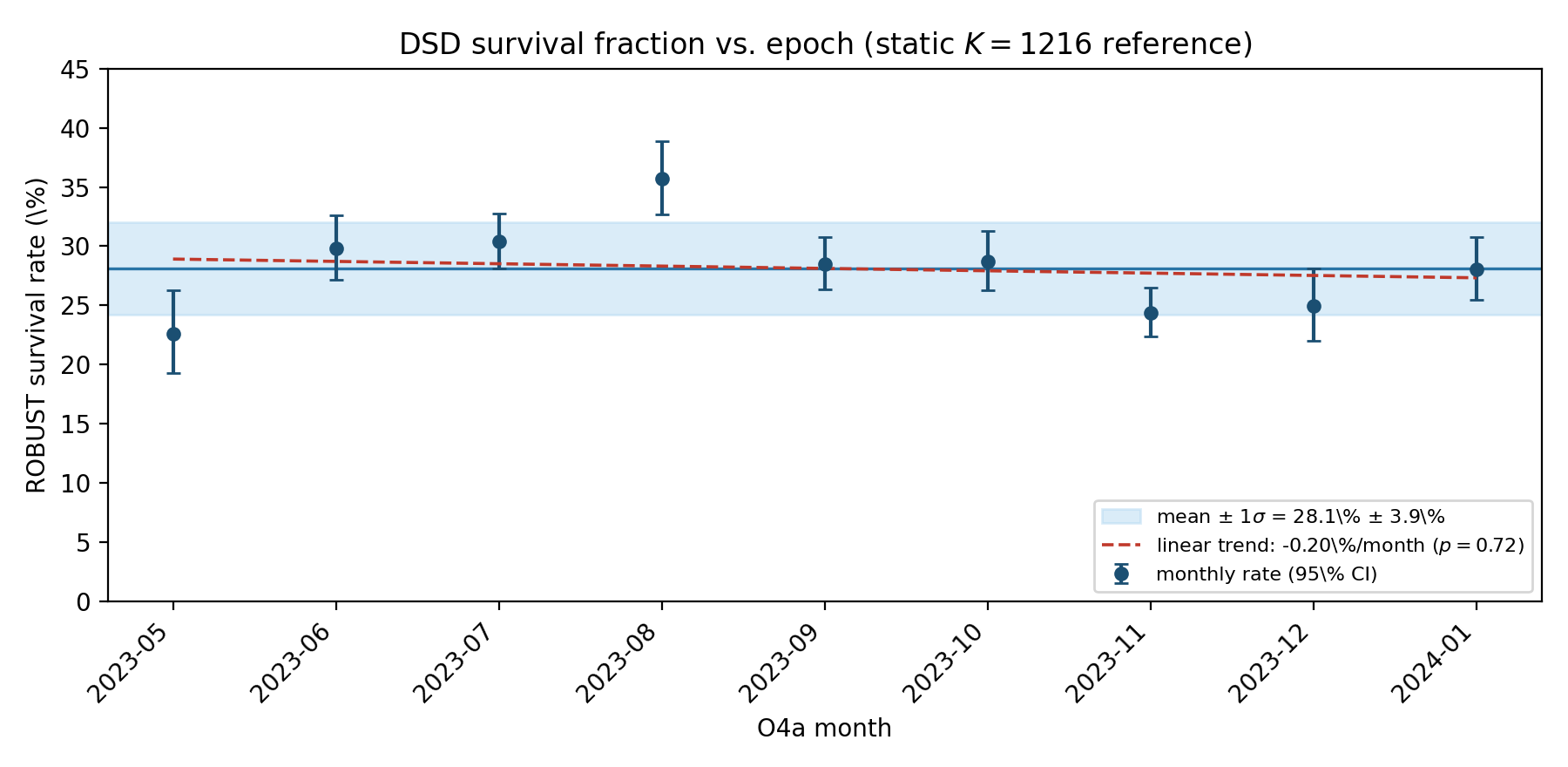}
    \caption{DSD survival fraction per calendar month, with Wilson 95\% binomial intervals, against the static $K=1216$ reference. Secular drift of the noise manifold against a fixed reference would produce a rising trend; the measured slope is slightly negative and consistent with zero ($-0.20\%$/month, $p=0.72$).}
    \label{fig:temporal_survival}
\end{figure*} It does not by itself establish that the survivors are a distinct morphological population, and the interpretation of \texttt{Family\_A} as an expression of manifold drift remains viable (Sec.~\ref{sec:diffusivity}).
    \item \textbf{Coincidence test: superseded statistic and residual limits of the replacement.} The embedding-similarity formulation of the coincidence stage was found to be unsuitable and has been replaced by the physical cross-correlation test of Sec.~\ref{sec:coincidence}. Two defects motivated this. (i) \emph{Chromatic-domain mismatch:} the partner spectrogram was encoded without the production colormap while the stored candidate vector used it, so every cross-detector similarity compared mismatched domains (the same failure mode as bug B-DSD-1, which had been fixed in the DSD path but not here). Re-measuring all $7{,}911$ affected candidates with the corrected encoding raises the mean from $0.328$ to $0.554$ and the maximum from $0.529$ to $0.942$. The code is corrected and the affected cache invalidated. The zero-coincidence outcome is unchanged by the correction, but the headroom to $\tau_{\mathrm{coh}}=0.975$ falls to $0.033$. (ii) \emph{Insufficient discriminating power:} even after correction, injected coincident signals reach only $\approx 0.9$ while the null extends to $\approx 0.94$, so the distributions overlap at any threshold. Consequently we do not quote the previously reported maximum similarity, which was a raw-domain artifact. The replacement statistic is validated ($1.00$ for an identical waveform, $0.043$ for independent noise) and has been applied to $8{,}749$ candidates ($84.4\%$ of the pool); the remaining $1{,}623$ lacked retrievable partner strain in the open archive. One residual limitation of the replacement should be noted: the per-event null uses only $4$--$8$ time shifts, which is why the threshold is set on the pooled null rather than per event; a per-candidate significance would require a denser shift ladder. The recovery efficiency is now measured rather than assumed --- $2{,}400$ dual-detector injections across ten morphologies, all reaching $\varepsilon_{\mathrm{coh}}=100\%$ at sufficient SNR (Table~\ref{tab:eps_coh}) --- but that efficiency is strongly morphology-dependent, saturating near $\mathrm{SNR} \approx 55$--$80$ for structured morphologies and only near $\mathrm{SNR} \approx 780$ for noise-like ones, so the null coincidence result is correspondingly weaker for incoherent broadband bursts. Note that the upper limits of Sec.~\ref{sec:results} do not depend on this stage: they bound instrumental morphologies per detector, as discussed there.
    \item \textbf{Centroid-based background reference in the diffusivity test.} The morphological diffusivity comparison (Sec.~\ref{sec:diffusivity}) uses the $K=1216$ $k$-means centroids as the background sample, because the raw pre-quantization background embeddings are not retained by the production pipeline. Centroid--centroid distances upper-bound the point-to-point spread of the underlying background population, so the test is indicative rather than a strict two-sample comparison of like objects; the reported conclusion accordingly rests on the distributional overlap and shape, not on the mean-distance ordering.
    \item \textbf{Taxonomy instability in small sessions.} Per-session clustering remains unstable for sessions with fewer than $\sim 100$ candidates (14 of 42 L1 and 19 of 42 H1 sessions fall below this cut); such sessions are excluded from the stability statistics of Sec.~\ref{sec:stability} and their per-session family assignments should not be over-interpreted, although their candidates fully participate in the global cross-session analysis.
    \item \textbf{Data-quality gating.} We gate only on \texttt{CBC\_CAT1}. Unmodeled-burst searches on the same O4a data additionally exclude CAT2 times \cite{lvk_burst_o4a, gwosc_o4a_release}; DANTE therefore analyzes a glitch-richer dataset than a burst analysis. As argued in Sec.~\ref{sec:scope} this is the appropriate choice for a glitch-discovery instrument --- CAT2 intervals are flagged for being glitch-rich, and excluding them would discard the object of study. We have quantified the effect on the denominator: intersecting our analyzed sessions with \texttt{\{DET\}\_BURST\_CAT2} removes a further $0.19$\,d from H1 ($0.13\%$ of $144.23$\,d) and $0.17$\,d from L1 ($0.12\%$ of $149.38$\,d), moving the limits from $R_{90} \le 5.83$ to $5.84\,\mathrm{yr}^{-1}$ (H1) and from $5.63$ to $5.64\,\mathrm{yr}^{-1}$ (L1) --- a shift below the precision at which we quote them. The remaining caveat is therefore not the livetime but the candidate pool: the survival fractions reported here are those of a CAT1-gated dataset and are not what a CAT2-gated analysis would obtain, since the excluded times are preferentially glitch-rich.
    \item \textbf{Threshold uncertainty, multiplicity, and the limits of FDR control.} The bootstrap confidence interval quantifies the sampling uncertainty of the $P_{99}$ flagging threshold, not the multiple-comparison inflation of the survivor \emph{count}. At the population level the excess is unambiguous: a nominal $P_{99}$ cut admits $\sim 1\%$ ($103.7$) of the $10{,}372$ candidates by chance under a pure-background hypothesis, against $2{,}937$ observed, a $280\sigma$ binomial excess. Per-candidate attribution is a different question, and we report explicitly that it cannot be settled with the present background sample. Applying Benjamini--Hochberg control at $\alpha = 0.01$ to empirical $p$-values computed against the native background returns zero discoveries --- but this is a resolution artifact, not a null result: with $5{,}000$ background samples per detector the smallest attainable $p$-value is $2 \times 10^{-4}$, whereas BH requires $p \leq \alpha/n = 9.6 \times 10^{-7}$ for any discovery at this pool size. The floor is $207\times$ too coarse, so the procedure is guaranteed to return zero for \emph{any} dataset under this configuration (Fig.~\ref{fig:fdr_resolution}). A meaningful per-candidate FDR treatment would require a background sample of order $10^{6}$.

\begin{figure}[htbp]
    \centering
    \includegraphics[width=\linewidth]{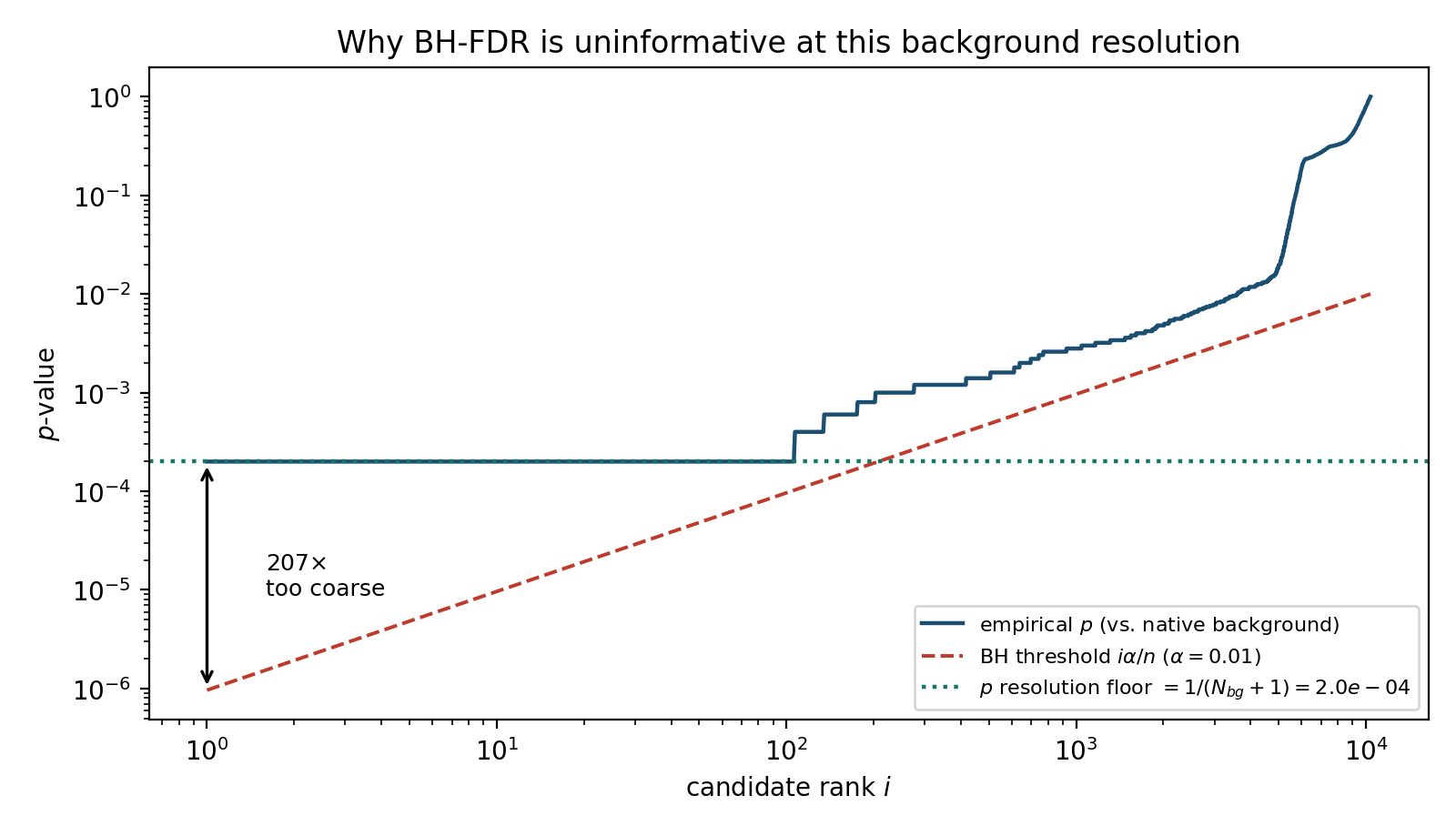}
    \caption{Benjamini--Hochberg control is resolution-limited here. Empirical $p$-values computed against the native background (blue) cannot fall below the floor $1/(N_{\mathrm{bg}}+1) = 2\times10^{-4}$ set by the $5{,}000$-sample background --- the first $\sim 100$ candidates all saturate it --- whereas BH requires $p \leq i\alpha/n$, which at rank $1$ demands $9.6\times10^{-7}$. The $207\times$ gap means no discovery is attainable for any dataset in this configuration, so the zero result carries no scientific content.}
    \label{fig:fdr_resolution}
\end{figure} We therefore rest the survivor-population claim on the collective excess and on the clustering topology, not on per-event significance.
    \item \textbf{Q-transform frequency span.} The pipeline requests a $20$--$2048$\,Hz Q-transform, but for $q \in [4, 64]$ over a $32$\,s window the transform that is actually produced stops at $\approx 1291$\,Hz; the requested upper bound is silently reduced. Patch-row to frequency conversions therefore overestimated frequency by $+12\%$ at the bottom of the grid and $+57\%$ at the top, and the per-event frequency bands recorded for the coincidence test are correspondingly too high (median $f_{\mathrm{hi}}$ $1862$\,Hz reported against $1226$\,Hz corrected). Because the band is applied identically to the on-source segment and to every time-shifted null, the coincidence comparison is internally consistent and the null result is unaffected. The constant is corrected in the released code.
\end{enumerate}

\section{Conclusions}
\label{sec:conclusions}

In this work, we deployed DANTE V3 as a detector-characterization instrument to survey the unmodeled \emph{instrumental} transient parameter space of early O4a. By processing over 10,000 multi-scale morphological candidates, we empirically demonstrated the extreme vulnerability of deep uncalibrated algorithms to non-stationary instrumental drift: $71.7\%$ of candidates are indistinguishable from the drifting noise manifold once re-scored against a native background index. The survivors form a single macro-cluster containing no compact recurring substructure; we also show that the diffusivity statistic previously used to characterize that cluster is confounded by its choice of background reference and by the threshold that defines the cluster, and we retire it accordingly. Applying the identical clustering procedure to unselected native background segments encoded through the same MIL path, we establish that the macro-cluster topology is not specific to the survivors --- pristine background is the most monolithic population of all --- so it cannot serve as a characterization of \texttt{Family\_A}. We regard this as a strengthening rather than a weakening of the survey's negative result: the pipeline does not merely fail to resolve a discrete recurring glitch family among the survivors, it demonstrably cannot resolve one for any population at this stage, which localizes the limitation in the morphological representation rather than in the data. One asymmetry does survive the test --- the DSD survivors are the only population that produces morphological isolates at all, three against zero in every control --- but at $p=0.12$ it does not reach significance and we report it as a falsifiable prediction for future runs rather than as a result. Through a formally calibrated physical environment monitoring veto, we anchored one of the two surviving isolated anomalies to an instrumental control coupling, and conservatively classified the other --- which shows no auxiliary coupling and no cross-detector counterpart, carries the highest anomaly score of the entire pool, and is one of only two candidates of $10{,}372$ outside the global macro-cluster --- as an uncatalogued instrumental morphological outlier. We stress that ``uncatalogued'' here refers to our own taxonomy: no Gravity Spy classification set covering O4a is publicly released, so we make no claim about this event's presence or absence in external glitch catalogs. We further show that the embedding-similarity formulation of the cross-detector coincidence test cannot separate an injected coincident signal from its own null, and replace it with a physical cross-correlation test over the light-travel lag window, validated at $1.00$ for an identical waveform against $0.043$ for independent noise; applied to $8{,}749$ candidates it finds no coincident event, with the on-source tail falling \emph{below} its own time-shifted null --- now a statement with demonstrated power rather than an empty one. The rate bounds reported here, $R_{90} \le 5.83\,\mathrm{yr}^{-1}$ (H1) and $R_{90} \le 5.63\,\mathrm{yr}^{-1}$ (L1), remain limits on novel instrumental glitch morphologies and carry no astrophysical interpretation. Together these results provide a blueprint for validating --- and for knowing the limits of --- unmodeled machine-learning discoveries in gravitational-wave detector characterization.

\begin{acknowledgments}
This research has made use of data or software obtained from the Gravitational Wave Open Science Center (gwosc.org), a service of LIGO Laboratory, the LIGO Scientific Collaboration, the Virgo Collaboration, and KAGRA. The auxiliary-channel veto makes direct use of the GWOSC O4 Auxiliary Channel Data Release (DOI: \href{https://doi.org/10.7935/kt51-6n86}{10.7935/kt51-6n86}) \cite{gwosc_aux_o4}; we acknowledge the invaluable contribution of the Physical Environment Monitoring (PEM) arrays and of the LIGO detector characterization groups that curated this release.
\end{acknowledgments}

\section*{Funding}
This research did not receive support from any organization.

\section*{Competing Interests}
The author declares no competing interests.

\section*{Data Availability}
The frozen analysis artifacts of this work (master taxonomy, disposition ledger, PEM family-wise calibrations, per-scale efficiency curves, and the final discovery report) are archived on Zenodo (DOI: \href{https://doi.org/10.5281/zenodo.21451803}{10.5281/zenodo.21451803}). All input strain and auxiliary-channel data are publicly available from the Gravitational Wave Open Science Center (gwosc.org), including the O4 Auxiliary Channel Data Release (DOI: \href{https://doi.org/10.7935/kt51-6n86}{10.7935/kt51-6n86}).

\section*{Code Availability}
The complete pipeline is open source at \url{https://github.com/lucacirfeta/dante-gravi-signal-ml} (Apache License 2.0). The exact version used for every number in this manuscript is pinned at tag \texttt{3.5.0}; all experimentally validated invariants of the pipeline are protected by the regression suite shipped in the same tag.

\section*{Author Contributions}
L.C. conceived the project, developed the pipeline, performed the analysis, and wrote the manuscript.

\bibliographystyle{apsrev4-2}
\bibliography{references}

\end{document}